\documentclass[preprint]{aastex}

\slugcomment{submitted to AJ}

\shorttitle{Spectral energy distributions of galaxies and AGN in the SPITZER
SWIRE Legacy survey}
\shortauthors{M.Rowan-Robinson, et al}

\begin{document}


\title{Spectral energy distributions and luminosities of galaxies and AGN in the SPITZER SWIRE Legacy Survey}


\author{Michael Rowan-Robinson$^1$, Tom Babbedge$^1$, Jason Surace$^2$, Dave Shupe$^2$,\\
Fan Fang$^2$, Carol Lonsdale$^{2,3}$, Gene Smith$^3$, Maria Polletta$^3$, Brian Siana$^3$,\\
Eduardo Gonzalez-Solares$^4$, Kevin Xu$^2$, Frazer Owen$^5$,\\
Payam Davoodi$^6$, Herve Dole$^2$, Donovan Domingue$^{2,11}$, Andreas Efstathiou$^7$, Duncan Farrah$^2$,\\
Matt Fox$^1$, Alberto Franceschini$^8$, Dave Frayer$^2$, Evanthia Hatziminaoglou$^9$,\\
Frank Masci$^2$, Glenn Morrison$^2$, Kirpal Nandra$^1$, Seb Oliver$^5$, Natalie Onyett$^5$,\\
 Deborah Padgett$^2$, Ismael Perez-Fournon$^9$, Steve Serjeant$^{10}$, Gordon Stacey$^{12}$, Mattia Vaccari$^1$\\  }
\affil{
$^1$ Astrophysics Group, Blackett Laboratory, Imperial College London,
Prince Consort Road, London SW7 2BZ, UK,\\
$^2$ Infrared Processing and Analysis center, California Institute of Technology 100-22, Pasadena, CA 91125, USA,\\
$^3$ Center for Astrophysics and Space Sciences, University of California, San Diego, La Jolla, CA 92093-0424, USA,\\
$^4$ Institute of Astronomy, Madingley Rd, Cambridge, CB3, UK,\\
$^5$ National Radio Astronomy Observatory, P.O.Box 0, Socorro, NM 87801, USA,\\
$^6$ Astronomy Centre, CPES, University of Sussex, Falmer, Brighton, BN1 9QJ, UK,\\
$^7$ Department of Computer Science and Engineering, Cyprus College, 6 Diogenes St, Engomi, 1516 Nicosia, Cyprus,\\
$^8$ Dipartimento di Astronomia, Universita di Padova, Vicolo Osservatorio 5, I-35122, Padova, Italy,\\
$^9$ Instituto Astrofisica de Canarias, Via Lactea, 38200 La Laguna, S$/$C de Tenerife, Spain,\\
$^{10}$ Centre for Astrophysics and Planetary Science,  School of Physical Sciences, University of Kent, Canterbury,
Kent CT2 7HR, UK.\\
$^{11}$ Department of Chemistry and Physics, Georgia College and State University, CBX 082, Milledgeville, GA 31061\\
$^{12}$ Department of Astronomy, Cornell University, 220 Space Science Building, Ithaca, NY 14853, USA\\
}



\begin{abstract}
We discuss optical associations, spectral energy distributions and photometric redshifts
for SWIRE sources in the ELAIS-N1 area and the Lockman Validation Field.
The band-merged IRAC (3.6, 4.5, 5.8 and 8.0 $\mu$m) and MIPS (24, 70, 160 $\mu$m) data
have been associated with optical UgriZ data from the INT Wide Field Survey in 
ELAIS-N1, and with our own optical Ugri data in Lockman-VF.  Criteria for eliminating
spurious infrared sources and for carrying out star-quasar-galaxy separation are
discussed, and statistics of the identification rate are given.
32 $\%$ of sources in the ELAIS-N1 field are found to be optically blank (to r = 23.5) and
16 $\%$ in Lockman-VF (to r = 25).

The spectral energy distributions of selected ELAIS sources in N1 detected by SWIRE,
most with spectroscopic redshifts,
are modelled in terms of a simple set of galaxy and quasar templates in the optical 
and near infrared, and with a set of dust emission templates (cirrus, M82 starburst,
Arp 220 starburst, and AGN dust torus) in the mid infrared.

The optical data, together with the IRAC 3.6 and 4.5 $\mu$m data, have been used
to determine photometric redshifts.  For galaxies with known spectroscopic redshifts
there is a notable improvement in the photometric redshift when the IRAC data are used,
with a reduction in the rms scatter from 10 $\%$ in (1+z) to 7 $\%$.  While further spectroscopic
data are needed to confirm this result, the prospect of determining good photometric 
redshifts for much of the SWIRE survey, expected to yield over 2 million extragalactic 
objects, is excellent.  Some modifications to the optical templates were required in
the previously uninvestigated wavelength region 2-5 $\mu$m.

The photometric redshifts are used to derive the 3.6 and 24 $\mu$m redshift distribution and to
compare this with the predictions of models.
For those sources with a clear mid infrared excess, relative to the galaxy starlight 
model used for the optical and near infrared,  
the mid and far infrared data are modelled in terms of the same dust emission templates
(cirrus, M82, Arp 220, AGN dust torus).   The proportions found of each template type
are: cirrus 31 $\%$, M82 29 $\%$, Arp 220 10 $\%$, AGN dust tori 29 $\%$.  
The distribution of the different
infrared sed types in the $L_{ir}/L_{opt}$ versus $L_{ir}$ plane, where $L_{ir}$ and $L_{opt}$
are the infrared and optical bolometric luminosities, is discussed.

There is an interesting population of luminous cool cirrus galaxies, with
$L_{ir} > L_{opt}$, implying a substantial dust optical depth.  Galaxies with
Arp220 like seds, of which there is a surprising preponderance compared with
pre-existing source-count models, tend to have high ratios of infrared to
optical bolometric luminosity, consistent with having very high extinction.
There is also a high proportion of galaxies whose mid-ir seds are fitted by an AGN dust 
torus template (29$\%$).  Of these only 8 $\%$ of these are Type 1 AGN according to the 
optical-nir template fitting while 25$\%$ are fitted with galaxy templates in the optical-nir 
and have $L_{ir} > L_{opt}$, so have to be Type 2 AGN.
The remainder have $L_{ir} < L_{opt}$ so can be Seyferts, in which the optical AGN fails to be
detected against the light of the host galaxy. The implied dust covering factor, $\geq 75\%$, is 
much higher than that inferred for bright optically selected quasars.

\end{abstract}




\section{Introduction}

Infrared wavelengths hold the key to understanding the evolution of galaxies.  From the spectrum
of the extragalactic background radiation ( Hauser and Dwek 2001) we see that much of the light
emitted by stars is absorbed by dust and reemitted at mid and  far infrared wavelengths.  Only
by understanding infrared extragalactic populations can we hope to get a reliable census of the
star formation history of galaxies and estimate the fraction of dust-obscured AGN.

IRAS 12-25-60-100 $\mu$m colour-colour diagrams proved remarkably effective in identifying
the main contributing components to the far infrared spectra of galaxies (Helou 1986, Rowan-Robinson and Crawford 1989). 
The key components were identified as infrared cirrus (emission from dust in the interstellar medium of the galaxy,
illuminated by old stars and by quiescent star formation),
starbursts (with the local starburst galaxy M82 as the prototype, generally associated with galaxy interactions), 
and AGN dust torus emission (predominantly at 3-30 $\mu$m in
the rest-frame).  Very high optical depth starbursts like Arp 220, probably associated with violent mergers, 
constituted a fourth component (Condon et al 1991, Rowan-Robinson and Efstathiou 1993).

The Infrared Space Observatory broadly confirmed the reality of these infrared galaxy populations.
Mid-infrared spectroscopy with ISO confirmed the distinct nature of these components and gave
details of the mid-infrared PAH features  
(Xu et al 1998, Lutz et al 1998, Helou et al 2000, Laurent et al 2000, Dale et al 2001, Lu et al 2003).
The ELAIS survey (Rowan-Robinson et al 2004) provides the largest sample of galaxies with ISO photometry at
6.7, 15, 90 and 175 $\mu$m.  Rowan-Robinson et al (2004) show optical and infrared spectral energy distributions
of a sample of ELAIS galaxies, modelled in terms of a simple set of 4 templates, for each of which
radiative transfer models exist.  We have used the same four templates to fit
the spectral energy distributions (seds) of SWIRE sources in this paper.

The SPITZER SWIRE Legacy Survey is expected to detect over 2 million infrared galaxies in 49 sq deg
of sky (Lonsdale et al 2003, 2004).  This will be especially powerful for searches for rare objects
and for systematic studies of the link between star formation history and large-scale stucture.
To date we have examined IRAC 3-8 $\mu$m data and MIPS 24-160 $\mu$m data in a 0.3 sq deg validation area
in our Lockman field (Lockman-VF), and in 6.5 sq deg of the ELAIS-N1 field.  Although this is only a 
small sample of the SWIRE survey ($\sim 13\%$), we can already make a quantitative survey of the galaxy
population.

In the ELAIS-N1 area we can make a direct comparison with the ELAIS ISO survey (Rowan-Robinson et al 2004).  
Sources in common 
will have been observed in a total of 10 infrared bands between 3 and 200 $\mu$m, 15, 90 and
175 $\mu$m, from ISO and the 7 SPITZER bands.  This allows us to carry out a detailed study of the spectral
energy distributions (seds) of these galaxies.  These studies validate our photometric redshift estimates 
and allow us to make estimates of the bolometric luminosity in different dust emission components.

We present here a preliminary analysis of these seds and the characteristic luminosities of
the infrared galaxies in the two samples of sky.
Polletta et al (2005 in prep) give a parallel discussion of the seds, redshifts and
nature of SWIRE sources, with an emphasis on colour-colour diagram
diagnostics.  Hatziminaoglou et al (2005) discuss SWIRE observations of SDSS quasars and Franceschini et al
(2005) discuss SWIRE observations of Chandra X-ray sources in N1.  Surace et al (2005 in prep) discuss the 
SWIRE IRAC (3-8 $\mu$m) source-counts, Shupe et al (2005) discuss the SWIRE 24 $\mu$m source-counts,
and Oliver et al (2004) discuss the 3.6 $\mu$m angular correlation function.

The SPITZER telescope and mission are described by Werner et al (2004), the IRAC instrument is described by
Fazio et al (2004), and the MIPS instrument by Rieke et al (2004).  Preliminary discussions of
SPITZER galaxy populations have been given by Chary et al (2004) for the GOODS-SV survey,
Eisenhardt et al (2004) for the IRAC Shallow Survey, Yan et al (2004) for the FLS Survey,
and Lonsdale et al (2004) for the SWIRE Survey.  The main thrust of these early discussions is the
capability of SPITZER to detect star-forming galaxies at z $>$ 1.  However these discussions were hampered 
by the absence of redshift information.  In this paper we use photometric redshifts to give a fuller
picture of the infrared galaxy populations.

A model with $\Omega_0 = 0.3, \lambda_0 = 0.7$ and a Hubble constant of 72 $km/s/Mpc$ is used throughout.

\section{Optical associations}

Our starting point is the band-merged SWIRE 3.6-24$\mu$m point-source catalogues (Surace et al 2004),
which consist of sources detected with an SNR of at least 5 in one or more IRAC bands and their 24 $\mu$m associations
with an SNR of at least 3. The astrometric accuracy of the IRAC sources appears to be very good, better than $\pm$0.5 arcsec 
(2-$\sigma$).
Aperture fluxes are used for 24 $\mu$m and fainter IRAC sources, but Kron fluxes, which give a better 
estimate of the flux from extended sources, have been used for the IRAC bands if $S(3.6)_{Kron} > $ 1 mJy.
Our goal is to define samples of good-quality SWIRE sources with reliable optical identifications and 
sufficient photometric bands for photometric redshift estimation.   
We search for optical associations of the SWIRE sources, using the band-merged WFS survey UgriZ data for the N1 area
(Babbedge et al 2005, in preparation), complete to r $\sim$ 23.5,  and our Ugri survey data for Lockman-VF 
(Siana et al 2005, in preparation), complete to r $\sim$ 25, using a search radius of 1.5".
Analysis of the identification rate as a function of search radius shows that an increasing fraction of
associations at radii larger than this are spurious (Surace et al 2004).  Where a SWIRE source picks up more than one optical
association, the association with the most optical bands is selected first (to select against a number of
spurious single band optical sources, and to favour brighter sources) and then the nearest if more than one have the same number of
optical bands.  Where two infrared sources closer than 0.5 arcsec acquire the same optical association we 
have kept only the nearer, since we suspect the majority of these close pairs may be processing artefacts.  
However there are some genuinely blended sources which will require a more sophisticated treatment.  After this
elimination of duplicates there are 16149 band-merged (IRAC+MIPS24) sources in the 0.3 sq deg Lockman area with optical associations
in at least one optical band,
and 204866 band-merged sources in the 6.5 sq deg of ELAIS-N1. 
Examination of band-merged sources with no optical association showed that some of these are spurious,
generated for example by bright stars (see discussion by Surace et al 2004).  The requirement that appears to 
eliminate spurious sources effectively and give a high reliability catalogue is that
the source be detected at both 3.6 and 4.5 $\mu$m at SNR $\ge$ 5 (Surace et al 2004).  This yields 96155 blank fields in N1 and
2991 in Lockman-VF, so that the identified sources represent 68 and 84 $\%$ of the total.  
Restricting to sources detected at 24 $\mu$m (38601 in ELAIS-N1), 35 $\%$ are optically blank (to r = 23.5).
For comparison Yan et al (2004) found 38 $\%$ of 24 $\mu$m sources in the FLS Survey have r $>$ 23.0 and
17 $\%$ have r $>$ 25.5.
SWIRE blank fields will be discussed further in a subsequent paper (Lonsdale et al 2004, in preparation).
Table 2 gives a summary of the statistics of the SWIRE catalogue in the ELAIS-N1 area.

Finding charts for ELAIS optical identifications and a summary of optical properties can be found in Gonzalez-Solares et al (2004)
and at http://www.ast.cam.ac.uk/~eglez/eid/

\section{Star-galaxy separation}

Optical objects classified as extended by Sextractor are assumed to be galaxies.  Those classified as
'stellar' may be stars, quasars or distant galaxies.  As an initial cut at removing the stars, we
removed objects with r $<$ 23 classified as stellar in r.  However some of these will be quasars, so we
defined an initial 'quasar' subset as (1) objects with an 8 $\mu$m/r-band colour cooler than a 2000 K
blackbody (due to presence of an AGN dust torus), or (2) objects whose UgriZ 3.6, 4.5 $\mu$m seds are
best fitted with an AGN rather than a galaxy template.

Figures 1 and 2 show the 3.6, r, i colour-colour diagrams for the 'stars' and 'quasars' defined in this
way, for tile 3-2 in N1 
(the SWIRE N1 data was analyzed in 0.8 sq deg tiles, and for some figures in this paper it
is clearer to show results only for a single tile).
Clearly the 'stars' 
still include a significant number of extragalactic objects, which we can recover by keeping those
above the straight line locus shown.  Similarly the 'quasars' still have a stellar contamination
which can be removed by omitting objects below the line shown.  To show how effective this
procedure is, Figure 3 shows the g-r-i colour diagram for our final stellar catalogue in N1, with almost
all objects following the stellar colour locus.  We find that 13 $\%$ of N1 sources and 6 $\%$
of Lockman-VF sources are stars.  
A similar star-galaxy separation criterion has been proposed by Eisenhardt et al (2004).

A few bright stars do not satisfy the stellar rejection criterion because of saturation in the
optical bands.  They are easily recognized as having r $<$ 16, stellar or saturated flags, photometric redshifts
close to zero and very poor $\chi^2$, and have been reclassified

\begin{figure}
\plotone{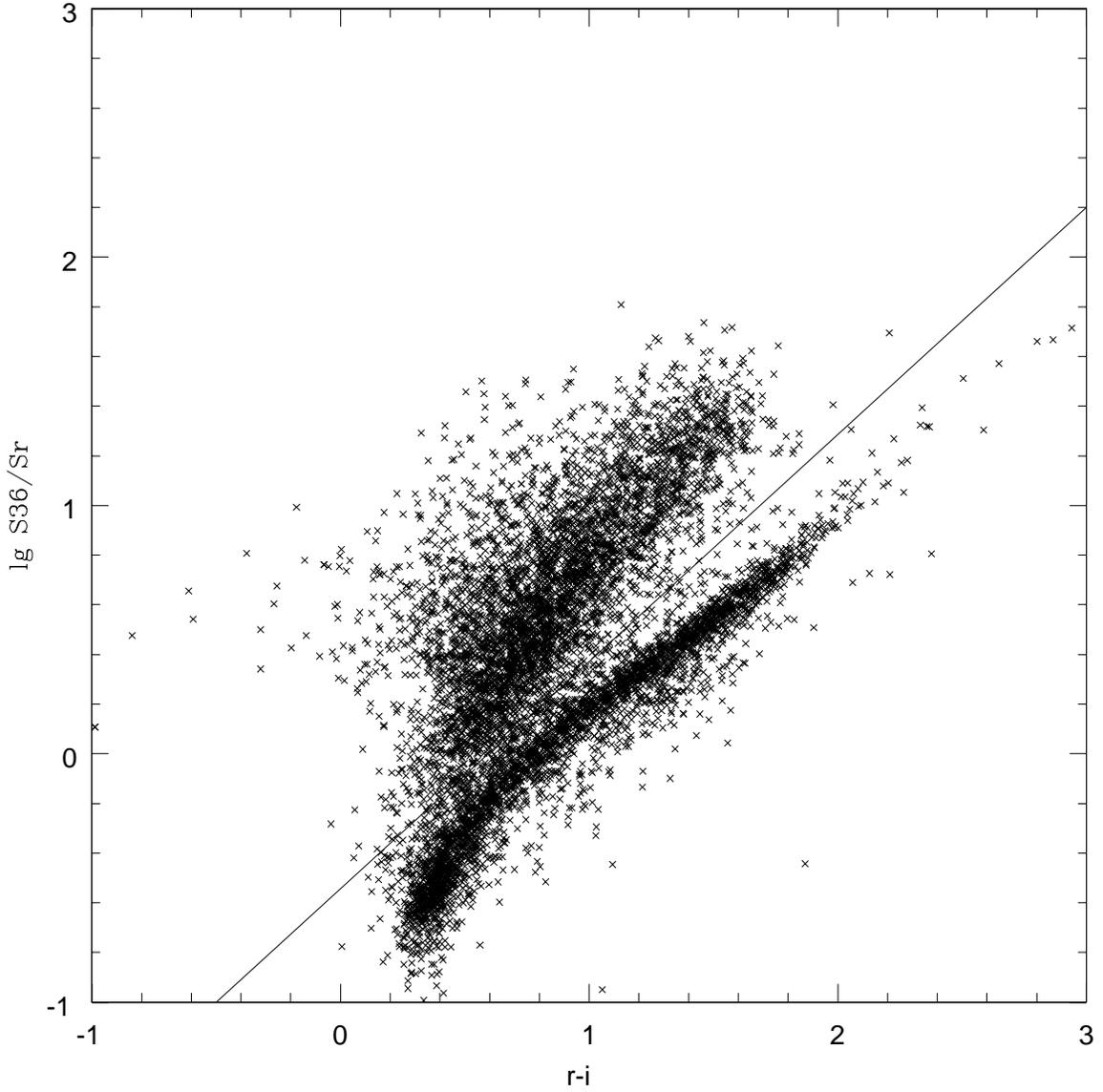}
\caption{3.6 $\mu$m-r-i colour-colour diagram for preliminary 'stars' 
catalogue in N1 (tile 3-2).
The line has equation 3.2(r-i+0.5) = 3.5 (log S(3.6)/S(r) +1).
}
\end{figure}

\begin{figure}
\plotone{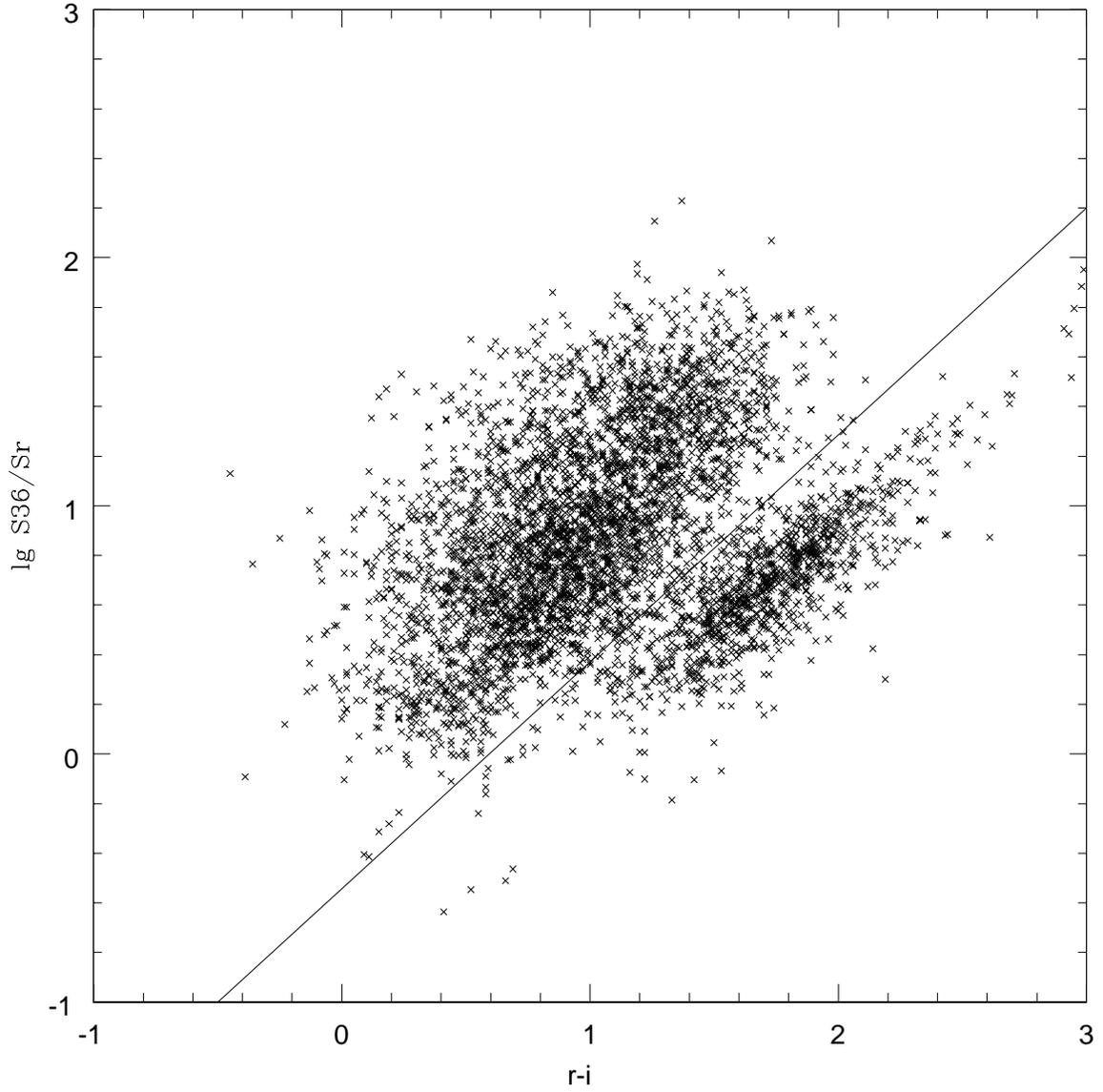}
\caption{3.6 $\mu$m-r-i colour-colour diagram for preliminary 'quasars' 
catalogue in N1 (tile 3-2).
}
\end{figure}

\begin{figure}
\caption{g-r-i colour-colour diagram for final stars catalogue in N1, with 17 $<$ r $<$ 22.
Fig 3 available at http://astro.ic.ac.uk/~mrr/
}
\end{figure}


\section{Spectral energy distributions}

To model the seds of SWIRE sources we have used an approach similar to that of Rowan-Robinson et al (2004) for
ELAIS sources.  The optical data and near infrared data to 4.5 $\mu$m are fitted with one of
the 8 optical galaxy templates used in the photometric redshift code of Rowan-Robinson et al
(2003), 6 galaxy templates (E, Sab, Sbc, Scd, Sdm, sb) and 2 AGN templates (Rowan-Robinson et 
al 2004).  The mid and far infrared data are fitted by a mixture of the 4 infrared templates
used by Rowan-Robinson (2001) in models for infrared and submillimetre source-counts: cirrus, M82 starburst,
Arp 220 starburst and AGN dust torus.  For each of these templates we have detailed
radiative transfer models (cirrus: Eftstathiou and Rowan-Robinson 2003, M82 and Arp 220 starbursts: Eftstathiou et al 
2000, AGN dust tori: Rowan-Robinson 1995, Efstathiou and Rowan-Robinson 1995).  The approach here is to try to
understand the overall SPITZER galaxy population. Better fits to individual galaxies could be obtained by
allowing variation of the many parameters in the radiative transfer codes from which these templates are taken and
this will be explored in subsequent work.  

With the first tranche of SWIRE data in the ELAIS-N1 region, we are able to fill in a major wavelength
gap in these seds and show for the first time high quality infrared seds from 0.36-175 $\mu$m, providing
a strong test of these models.  In figures 4-6 we show sed fits for 30 sources with spectroscopic redshifts which are detected in
most of the ISO and SWIRE bands (including MIPS 70 and 160 $\mu$m).
The ELAIS Final Catalogue was associated with the SWIRE 3.6-24 $\mu$m catalogue using a search radius
of 2.5 arcsec.  The MIPS 70 and 160 $\mu$m fluxes were extracted from mosaiced maps by Donovan Domingue using the Starfinder code (Diolaiti et al 2000).  
70 and 160 $\mu$m source-lists were associated with the SWIRE sources with 24 $\mu$m detections using a search radius of 7 arcsec
at 70 $\mu$m and 14 arcsec at 160 $\mu$m.  Histograms of source separation show an increasing fraction of
spurious associations at radii greater than these. Where 70 or 160 $\mu$ sources acquire more than one 24 $\mu$m
association, the flux was always assigned to the brighter 24 $\mu$m source.  In some cases, especially at 160 $\mu$m,
the flux should probably be assigned to more than one shorter wavelength source, so the 160 $\mu$m flux
will be overestimated.  For a few sources which are extended at 70 or 160 $\mu$m, these fluxes may be underestimated.
916 sources were associated in this way with 70 or 160 $\mu$m detections, with about 35 $\%$ of these being ELAIS sources, within the region of overlap of the two surveys.
The numbers of sources in the fully bandmerged 3.6-175 $\mu$m SWIRE+ELAIS catalogue
with 10, 9, 8 or 7 infrared bands out of 3.6, 4.5, 5.8, 8.0, 15.0, 24.0 70, 90, 160, 175 $\mu$m is 14, 44, 93, 332, 
respectively.  

Figure 4 shows seds for 10 galaxies with spectroscopic redshifts and which are detected 
at UgriZ, JHK (primarily from 2MASS), 3.6, 4.5, 5.8 8.0, 15, 24, 70, 90, 160
and 175 $\mu$m, and which were characterized by Rowan-Robinson et al (2004) as cirrus galaxies, ie normal spirals
with mid and far infrared emission from interstellar dust illuminated by the general stellar radiation field.  

\begin{figure}
\plotone{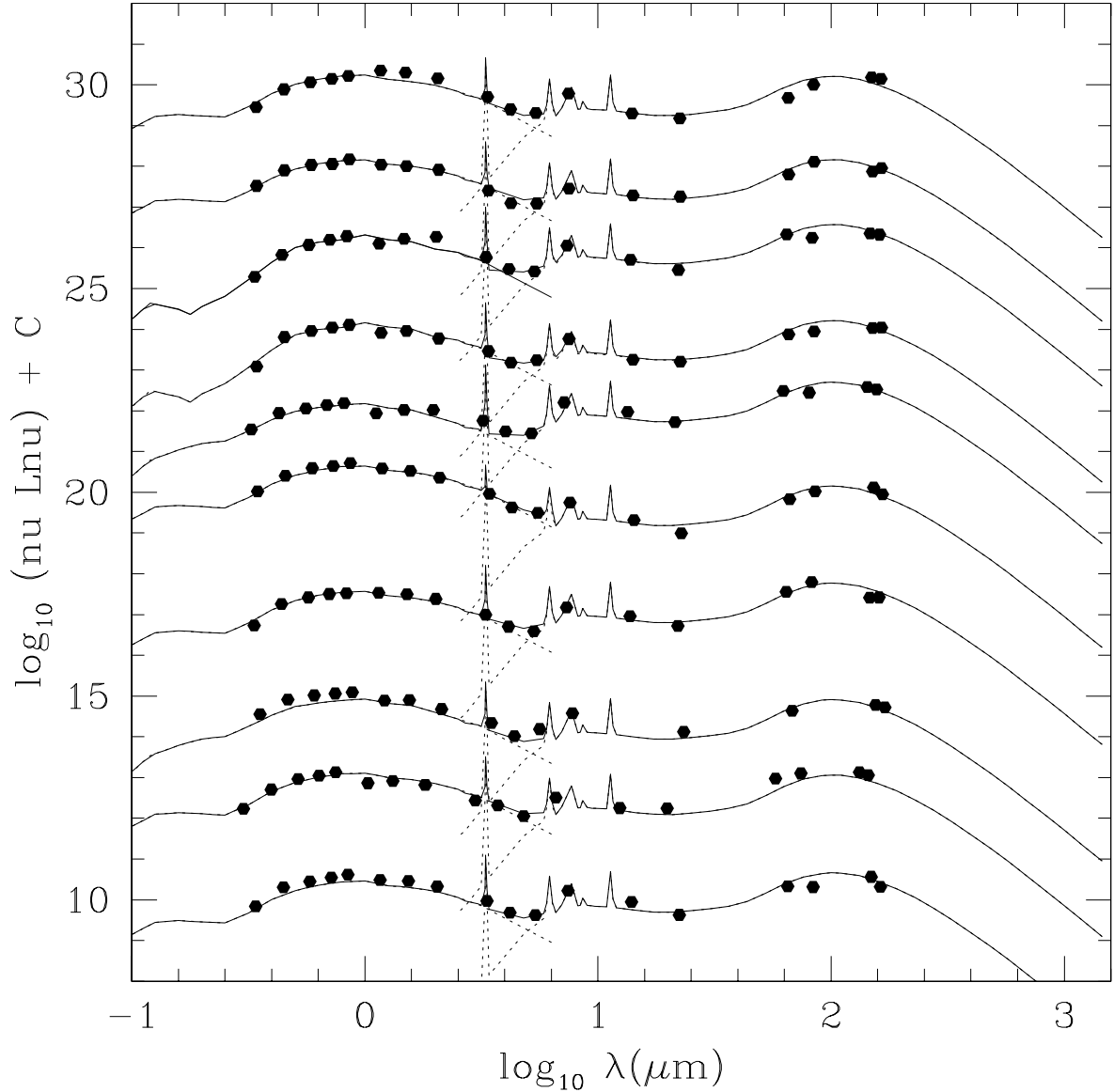}
\caption{Spectral energy distributions for 10 galaxies with spectroscopic redshift from the ELAIS survey, 
characterized by Rowan-Robinson et al (2004) as cirrus galaxies, with 10 infrared spectral bands at 
$\lambda > 3 \mu$m.  The SPITZER data fit well onto the 
cirrus template model of Rowan-Robinson (2001, see also Efstathiou and Rowan-Robinson 2003),
and there is good agreement between the ISO and SPITZER fluxes.
Dotted curves show the optical-nir and cirrus templates, solid curve the total predicted sed.
Data are the INT Wide Field Survey UgriZ, JHK (generally 2MASS), ISO 15,90, 175 $\mu$m,
IRAC 3.6, 4.5, 5.8, 8.0 $\mu$m, and MIPS 24, 70, 160 $\mu$m.  
Error bars (typically $\pm$0.04 dex, corresponding to a 10$\%$ photometric
uncertainty) are smaller than the size of the symbols.
Details of the sources in Figs 4-6 are given in Table 1, sequentially starting from the bottom of Fig 4.}
\end{figure}

\begin{figure}
\plotone{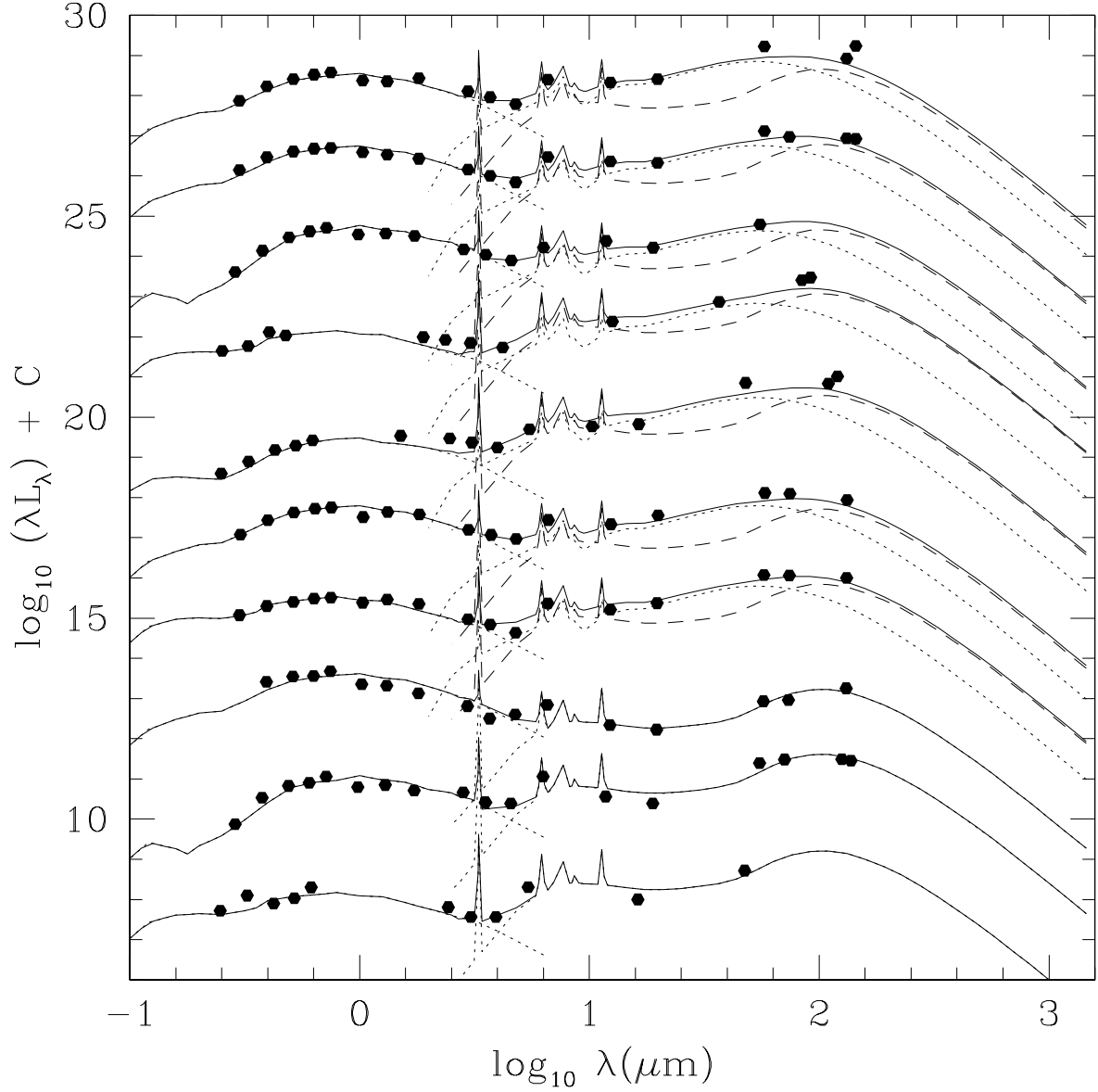}
\caption{Spectral energy distributions for 10 galaxies, 9 with spectroscopic redshifts, from the ELAIS survey.
3 are luminous cirrus galaxies, while 7 are fitted by a combination of an M82 starburst and cirrus.
The optical-nir and M82 templates are shown as dotted curves, the cirrus template as a broken curve,
and the total predicted sed as a solid curve.
}
\end{figure}

\begin{figure}
\plotone{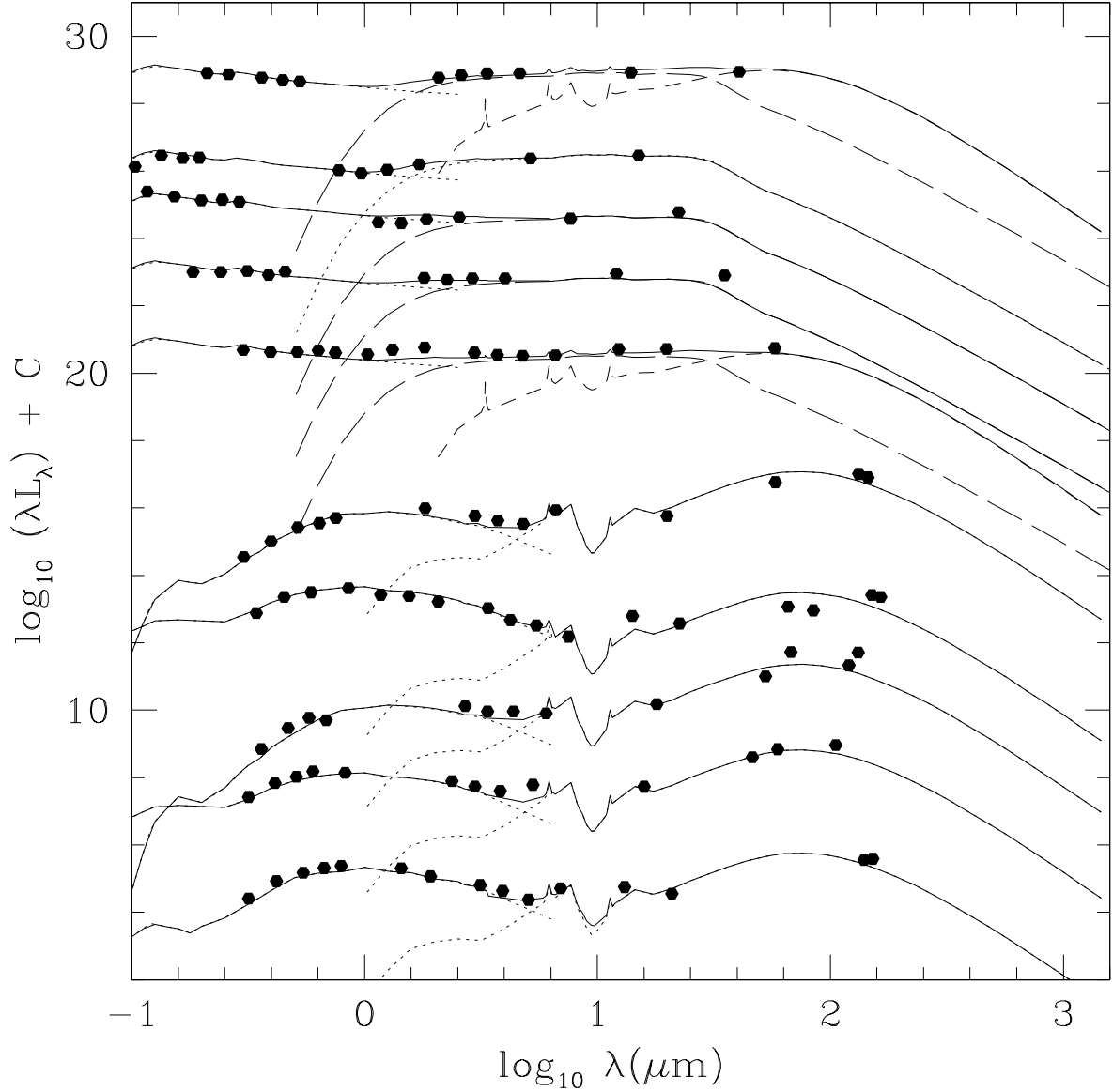}
\caption{Spectral energy distributions for 5 galaxies from the ELAIS survey
characterised as Arp 220 starbursts,
3 with spectroscopic redshifts,
and 5 quasars with spectroscopic redshifts, with infrared seds modelled with 
an AGN dust torus and, in two cases, an additional M82 starburst component.
The optical-nir and Arp 220 templates are shown as dotted curves, the AGN dust torus as a long-dashed
curve, the M82 starburst as a short-dashed curve, and the total predicted sed as a solid curve.
}
\end{figure}


Table 1 gives the parameters of the model fits: SPITZER RA and dec, ELAIS name, redshift (bracketed if
photometric), $n_{sed}$ (infrared sed type: 1 = cirrus, 2 = M82 starburst, 3 = Arp220 starburst,
4= AGN dust torus), $\log_{10}$ bolometric infrared luminosity $L_{ir}$, the vertical offset constant C
used in plotting the seds, the optical-nir sed type $n_{typ}$ (1 = E, 2 = Sab, 3 = Sbc, 4 = Scd, 5 = Sdm,
6 = sb, 7 and 8 = QSOs, see Rowan-Robinson 2003), $A_V$, $\log_{10}$ bolometric optical luminosity $L_{opt}$.
The data at wavelengths $<$ 5 $\mu$m are fitted with the optical galaxy templates used by
Rowan-Robinson et al (2004) in their photometric redshift estimation (see section 5).  The 
spectroscopic redshifts are in the range 0.03-0.27.  The cirrus template
used at wavelengths $>$ 5 $\mu$m is taken from the sequence of radiative transfer models by Efstathiou and Rowan-Robinson
(2003).  The SPITZER data fit remarkably well onto the models determined solely from ISO, optical data
and JHK data, demonstrating the consistency of the IRAC and MIPS calibrations. 
It is remarkable how well a
single cirrus template fits these relatively quiescent galaxies ($L_{ir} < 2.10^{11} L_{\odot}$).  A few galaxies (not shown here)
will require a cooler cirrus template.
There is a reasonable consistency
between the ISO 90 and 175 $\mu$m data and the MIPS 70 and 160 $\mu$m data.  For example for 
cirrus sources in common
it is found that $log_{10} (S(90)/S(70)) = 0.27 \pm 0.17$ and $log_{10} (S(175)/S(160)) = 0.02 \pm 0.12$, so the
consistency of both ISO and SPITZER calibrations is supported.

Figure 5 shows seds for 10 ELAIS-N1 galaxies which are characterized as higher-luminosity cirrus galaxies
or M82  starbursts,
all except one detected in at least two of the bands 15, 90 and 175 $\mu$m.  The M82 template
is taken from the sequence of radiative transfer models by Efstathiou et al (2000).   The galaxies have spectroscopic redshifts in the
 range 0.2-0.9 .  The three 'cirrus'
 galaxies are very luminous, with $L_{ir} > 3.10^{11} L_{\odot}$ and have $L_{ir}$ significantly larger
 than $L_{opt}$, implying a substantial optical depth in dust.  One (241.67583+55.83316) is ultraluminous
 ($L_{ir} = 2.4.10^{12} L_{\odot}$), based on a photometric redshift of 0.479.  
 Mid infrared spectroscopy
 and submillimetre photometry are needed to confirm the existence of this class of cool luminous
 galaxies, whose existence is predicted by the counts model of Rowan-Robinson (2001) and for which direct
 evidence is presented by Efstathiou and Rowan-Robinson (2003), Chapman
et al (2003), Kaviani et al (2003) and Rowan-Robinson et al (2004).  These are discussed further in section 7. 
 
Figure 6 shows seds for 10 ELAIS-N1 galaxies, comprising 5 Arp 220  (higher optical depth) starbursts, and 5 AGN dust tori.
Two of the latter also show evidence of a starburst component at longer wavelengths.
The Arp 220 template is taken from the sequence of radiative transfer models by Efstathiou et al (2000) and
the AGN dust torus template is taken from the radiative transfer model of Rowan-Robinson (1995).
 The objects have spectroscopic redshifts in the range 0.06-3.6 (two of the Arp 220 objects have photometric redshifts only).
The fits for 160901+541808 and 161005.8+541029 are worse than for other objects plotted at $\lambda > 60 \mu$m.

The striking features that emerge from this modelling are (1) that the seds can at least broadly be understood in terms of 
a small number of infrared templates, (2) the emergence of the class of luminous cool
galaxies with $L_{ir} > L_{opt}$, (3) the significant incidence of Arp 220 type seds in the survey, which was not predicted by 
source count models (eg Rowan-Robinson 2001).  Radiative transfer codes have several parameters and even better fits could 
be found by using the full range of models of Efstathiou et al (2000), Eftstathiou and Rowan-Robinson (2003).  This
will be explored by Efstathiou et al (2004, in preparation).

In fitting these seds, and in the photometric redshift analysis described in the next section, it became 
apparent that the predictions of the optical-nir templates used by Rowan-Robinson (2003) at wavelengths
2-5 $\mu$m, where they had not previously been tested, were too low.  Figures 7 and 8 show a comparison of 
data for galaxies with spectroscopic redshifts and which are best fitted by E and Scd templates, respectively.
Stellar synthesis fits to these seds (eg Babbedge et al 2004) will require a larger population of old low-mass
stars. 

\begin{figure}
\plotone{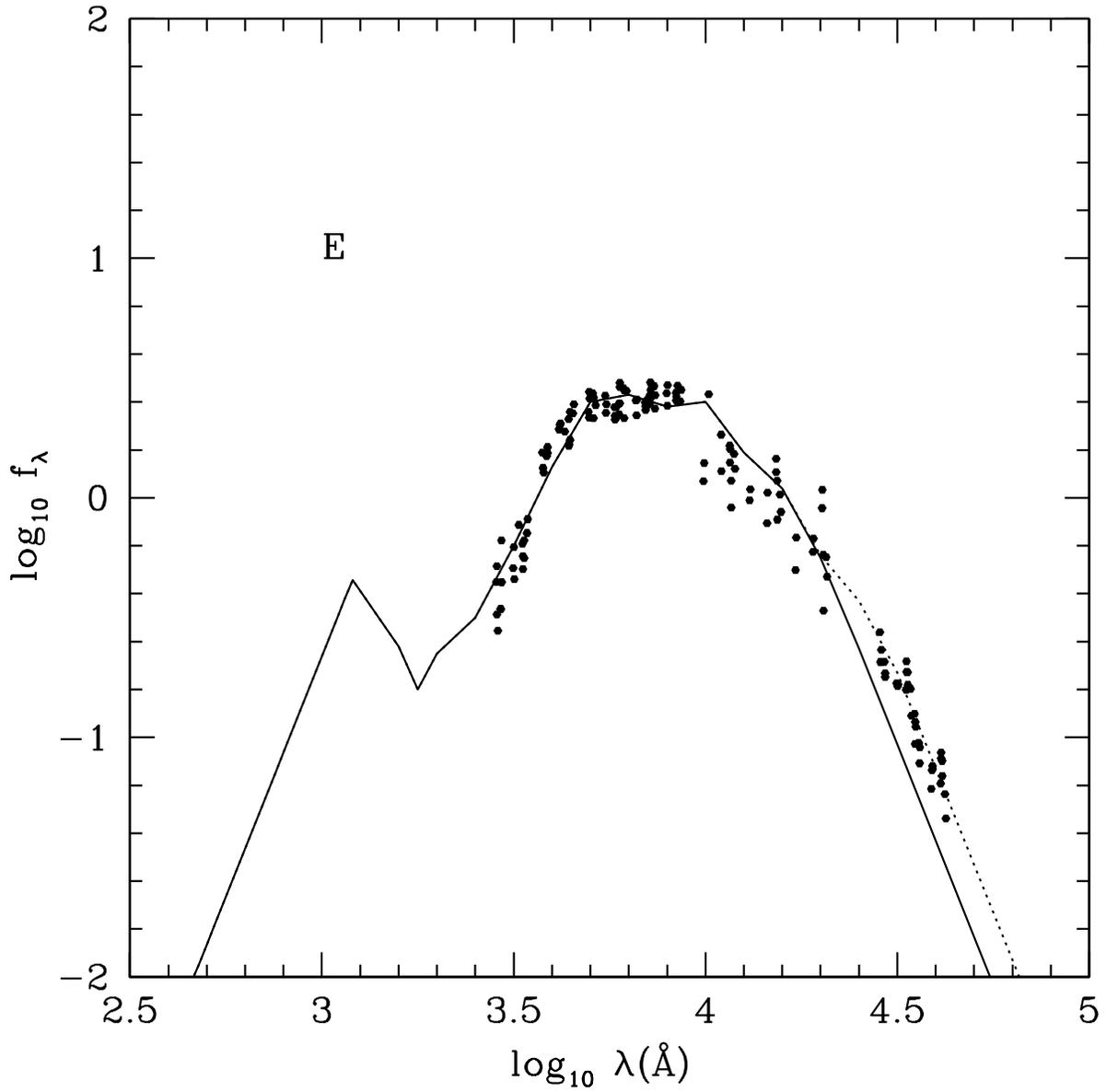}
\caption{Comparison of normalized photometric data for SWIRE-ELAIS galaxies with spectroscopic redshifts
whose seds are classified as E
with optical-nir sed template for elliptical galaxies used in the photometric redshift code, illustrating the need
to refine the templates at wavelengths $> 2 \mu$m.
}
\end{figure}

\begin{figure}
\plotone{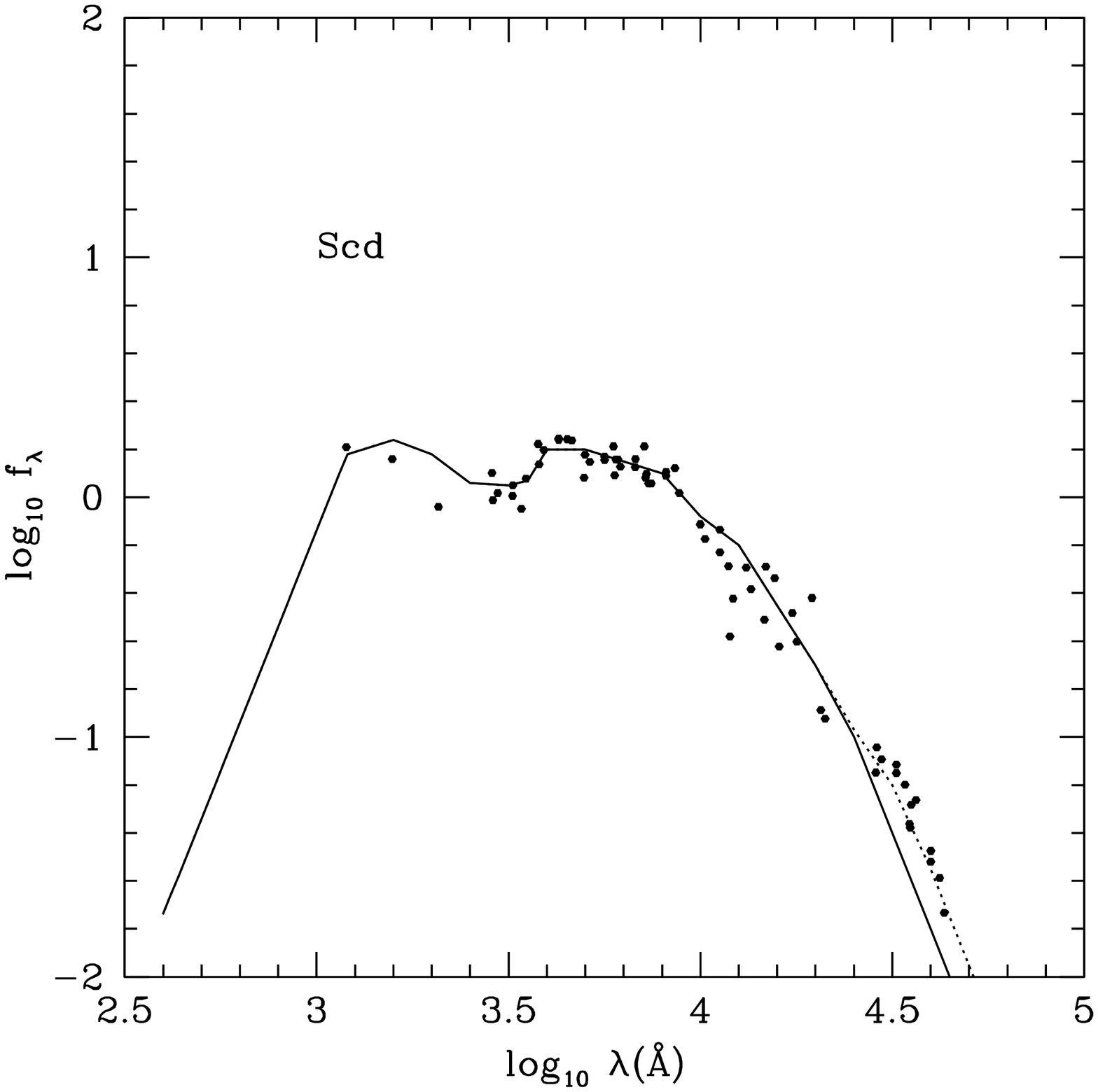}
\caption{Comparison of normalized photometric data for SWIRE-ELAIS galaxies with spectroscopic redshifts
whose seds are classified as Scd
with optical-nir sed template for Scd galaxies used in the photometric redshift code, illustrating the need
to refine the templates at wavelengths $> 2 \mu$m.
}
\end{figure}


\section{Photometric redshifts in ELAIS-N1 and Lockman-VF}

We have been able to carry out a photometric redshift analysis for the two SWIRE samples, the 0.3 sq deg area of 
Lockman-VF, in which we have moderately deep Ugri optical data (to r $\sim$ 25), and the  6.5 sq deg of the N1 area, in which we
have the re-reduced WFS UgriZ data (to r $\sim$ 23.5).  Full analyses of the photometric data in these fields 
will be reported by Siana  et al (2005, in preparation) and Babbedge et al (2005, in preparation).
Figure 9 shows a plot of $log_{10} (S(3.6))$ versus r for the Lockman-VF field, which illustrates
that the optical selection starts to produce serious incompleteness below S(3.6) $\sim$ 40 $\mu$Jy for an optical
survey complete to r = 23.5 (as in ELAIS-N1), and below S(3.6) $\sim$ 10 $\mu$Jy for an optical survey complete 
to r = 25 (as in the Lockman-VF).  Figure 10 shows $log_{10} (S(24))$ versus r for a 0.8 sq degree sub-area of N1,
with symbols and colours denoting the different infrared template types, discussed in section 7. 
Optical incompleteness becomes serious for 24 $\mu$m sources below 200 $\mu$Jy for a survey complete to r = 23.5.

\begin{figure}
\plotone{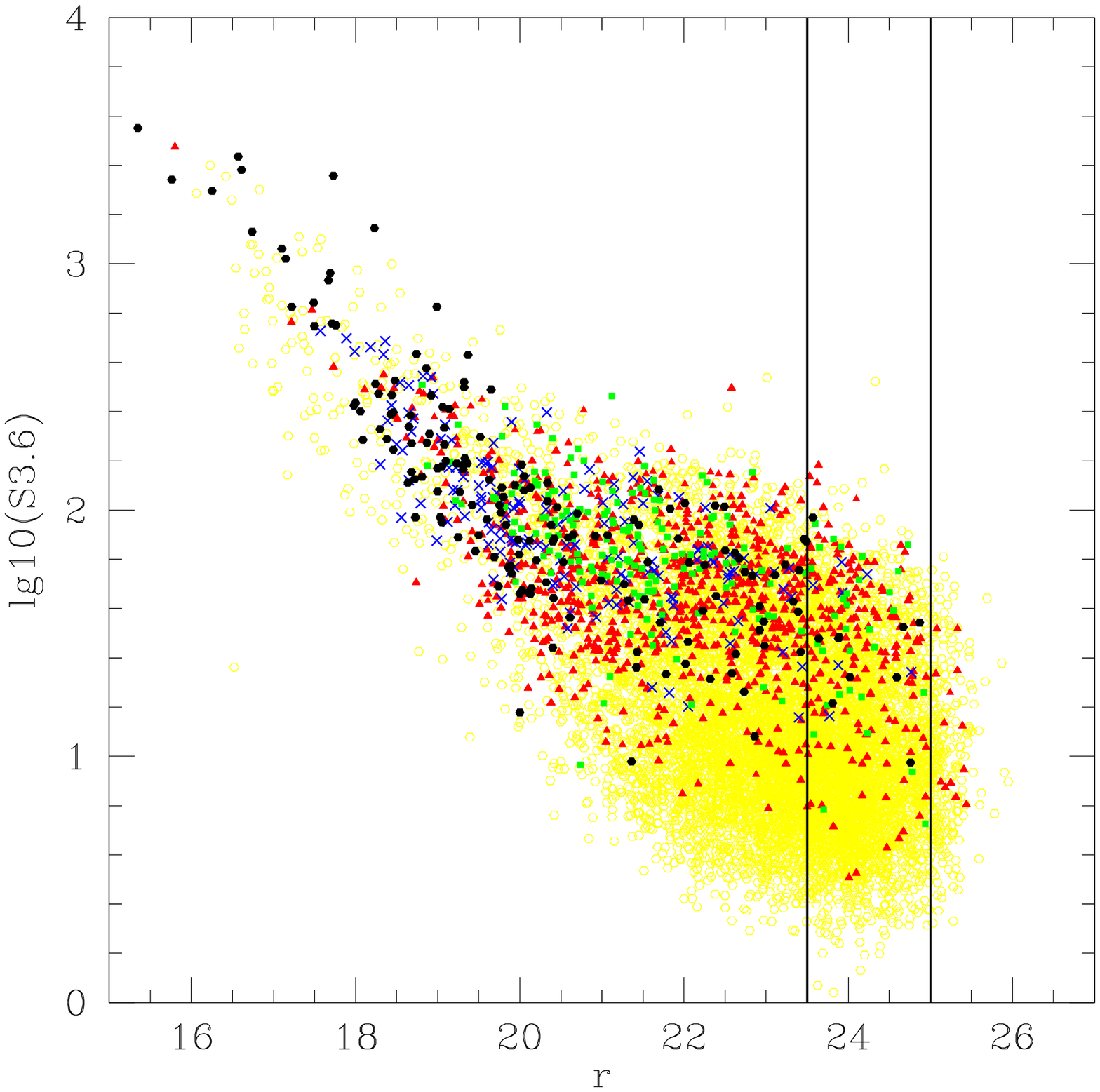}
\caption{ $log_{10} S(3.6)$ versus r for identified SWIRE sources in Lockman-VF, 
with different symbols for infrared template types:
filled circles (black): cirrus, filled triangles (red): M82 starburst, filled squares (green): Arp220,
crosses (blue): AGN dust tori.
Sources with no infrared excess or excess in only a single band are shown as yellow open circles.
Vertical lines indicate the optical completeness of the WFS and Lockman-VF surveys.
}
\end{figure}

\begin{figure}
\plotone{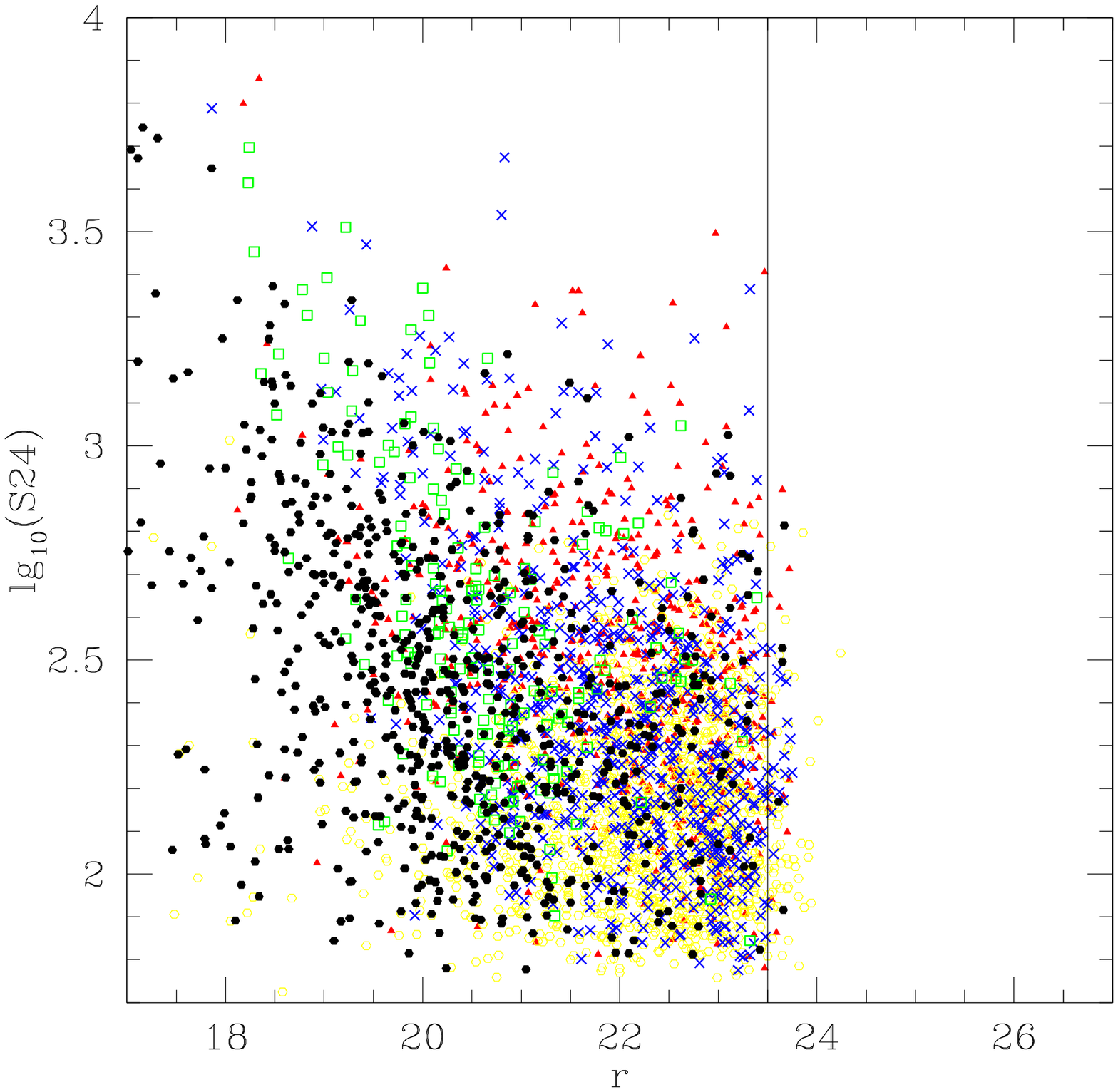}
\caption{$log_{10} S(24)$ versus r for identified SWIRE sources in N1 (tile 2-2), with different symbols for
infrared template types:
filled circles (black): cirrus, filled triangles (red): M82 starburst, open squares (green): Arp220,
crosses (blue): AGN dust tori.
}
\end{figure}

In this paper we have analyzed the catalogues of SWIRE sources with optical associations 
with the code of Rowan-Robinson (2003) which uses 6 optical 
galaxy templates and 2 AGN templates (as described in Rowan-Robinson et al 2004).  
The galaxy templates were based on empirical seds by Yoshii and Takahara (1988) and Calzetti and Kinney (1992) but have been subsequently modelled with
a stellar synthesis code by Babbedge et al (2004).  Parallel work is being carried out with IMPz, a more detailed
version of this code described by Babbedge et al 2004 (Babbedge et al 2005, in preparation), and with a modified version of the
hyper-z code (Bolzonella et al 2000) developed by M.Polletta (Polletta et al 2005 in preparation).  Agreement between 
these three approaches applied to ELAIS data was found to be good by Babbedge et al (2004).

We set a further condition on the optically associated SWIRE sources for entry to the photometric redshift code, 
namely that the source be detected at 3.6 $\mu$m, in the r-band, and in at least two other bands.  Rowan-Robinson
(2003) found that four bands was the minimum for effective photometric redshift estimation.
This reduced the sample of optically identified SWIRE sources (after elimination of stars) to
133645 in N1 and 12664 in Lockman-VF.   
For only 6 $\%$ of these 4-band optically associated extragalactic SWIRE sources in N1 does the code fail to 
 produce a redshift, with the reduced $\chi^2 >$ 10, and 12 $\%$ in Lockman-VF.  To quantify the goodness of fit,
figure 14 shows the histogram of reduced $\chi^2$
for N1 galaxies, which is consistent with the templates being used providing a reasonable fit to the optical-near-infrared data.

For a sample of the brighter Lockman-VF galaxies (the WIYN survey, Owen et al 2004, in preparation), and 
for 25$\%$ of the galaxies in N1 which are common to the ELAIS survey (Perez-Fournon et al 2004, Serjeant et al 2004, 
summarized in Rowan-Robinson et al 2004), we have spectroscopic redshifts 
and can make a direct comparison with the derived photometric redshifts.
An important innovation offered by the SPITZER data is that we can use the 3.6 and 4.5 $\mu$m fluxes 
in the photometric redshift fits.  Dust emission in cirrus and starburst galaxies generally starts to be a significant contributor to
the total flux at around 5 $\mu$m in the rest frame.
We find that inclusion of extinction, characterized by $A_V$, is essential for a sensible
redshift solution for SWIRE galaxies, some of which appear to be dusty, with $A_V$ ranging up to 2.  
We also find it necessary to allow modest extinction for the QSOs.  We restricted $A_V$ for QSOs to $\le$ 0.3,
since allowing a greater extinction range lets in many more aliases for normal galaxies.
This may result in some heavily reddened quasars not achieving a correct fit.
Figure 11 shows a comparison of
spectroscopic and photometric redshifts for 120 galaxies in the SWIRE-N1 area, derived from UgriZJHK data plus
SWIRE 3.6 and 4.5 $\mu$m fluxes.  The rms deviation of $log_{10} (1+z)$ for galaxies is 6.9 $\%$, a significant 
improvement.  For comparison typical rms values found by Rowan-Robinson (2003) and Rowan-Robinson et al (2004)
from UgriZ, JHK data alone were 10 $\%$.  Figure 12 shows a comparison of
spectroscopic and photometric redshifts for 289 galaxies in the WIYN survey in Lockman-VF derived from Ugri
photometry, plus 3.6 and 4.5 $\mu$m fluxes.  The rms deviation of $log_{10} (1+z)$ is 8.4 $\%$, again a
significant improvement over determinations from Ugri data alone   
(the corresponding plot when 3.6 and 4.5 $\mu$m data are not used is shown in Fig 13).
Use of IRAC data in photometric redshift determination has also reduced the number of outliers.
Redshift estimates for AGN can be affected by optical variability since optical photometry for
different bands may have been taken at different epochs (Afonso-Luis et al 2004).
A fuller discussion of photometric
redshifts of SWIRE galaxies will be given by Babbedge et al (2005, in preparation).

The distribution of optical-ir sed types for the 126193 galaxies in N1 for which we have photometric redshift
estimates is given in Table 2.  The number of ellipticals my be inflated by objects similar to Arp 220, whose
optical sed looks very like an elliptical because of the heavy obscuration of the newly-forming stars.

\begin{figure}
\plotone{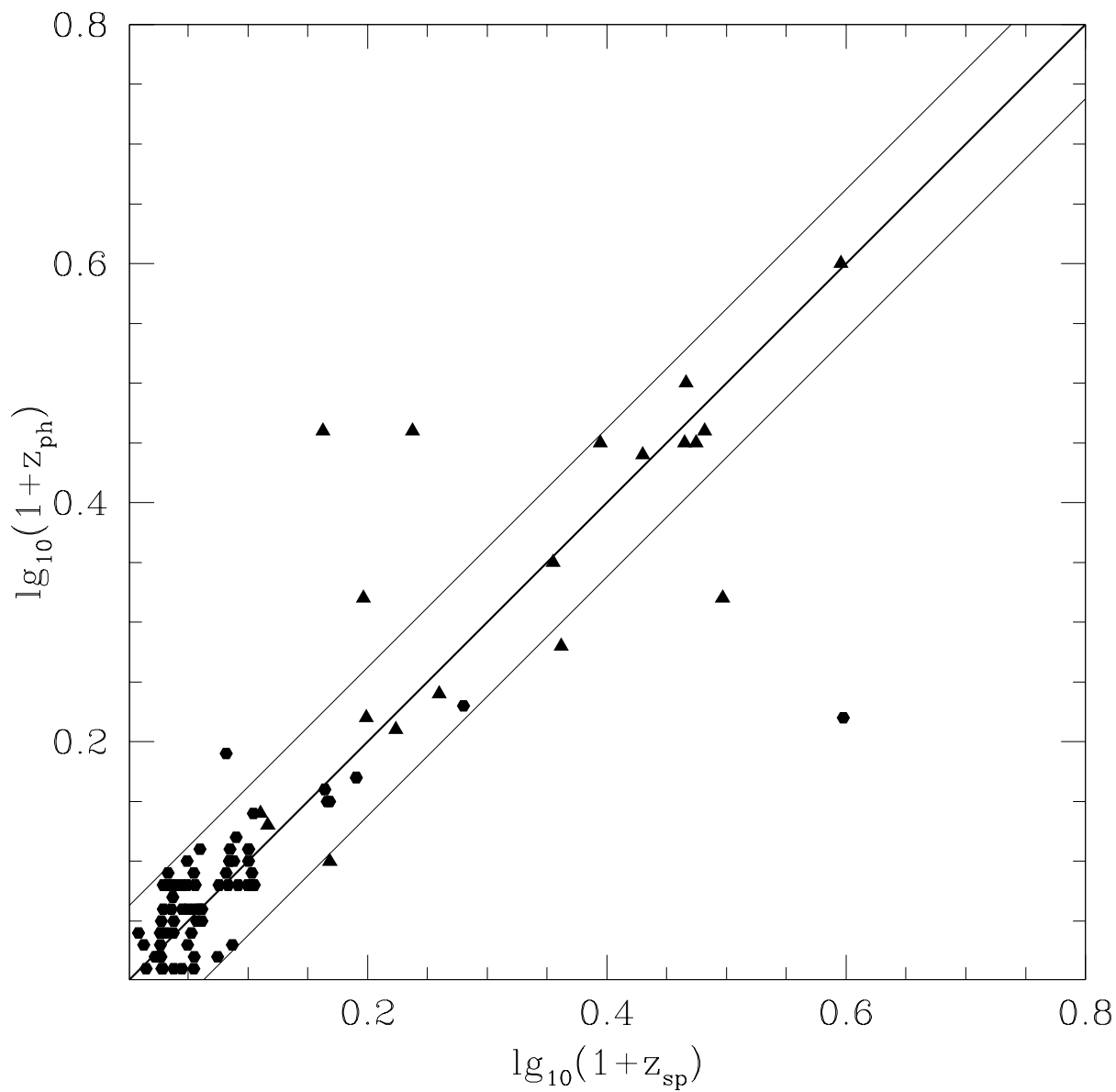}
\caption{Photometric redshifts in ELAIS-N1 using UgriZ, JHK, and SWIRE 3.6 and 
4.5 $\mu$m data, versus spectroscopic redshift.
Filled circles: galaxies, filled triangles: QSOs.
The $\pm 2 \sigma$ loci are indicated.
Spectroscopic redshifts are from Perez-Fournon et al (2004, in preparation), Serjeant et (2004) (see also Rowan-Robinson
et al 2004), Hatziminaoglou et al 2004.
}
\end{figure}

\begin{figure}
\plotone{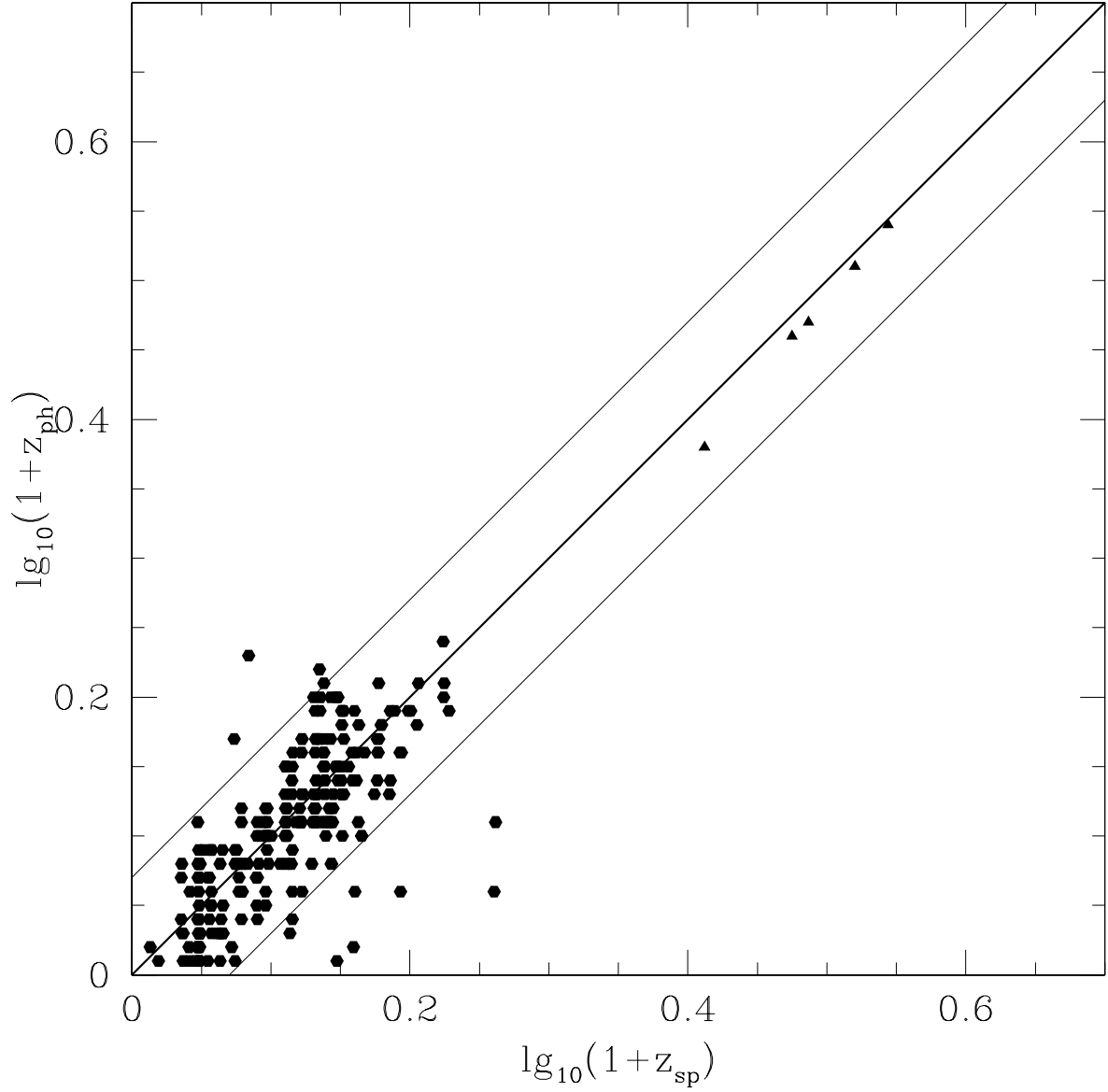}
\caption{Photometric redshifts for the WIYN survey using Ugri and SWIRE 
3.6 and 4.5 $\mu$m data, versus spectroscopic redshift.
Filled circles: galaxies, filled triangles: QSOs.
The $\pm 2 \sigma$ loci are indicated.
}
\end{figure}

\begin{figure}
\plotone{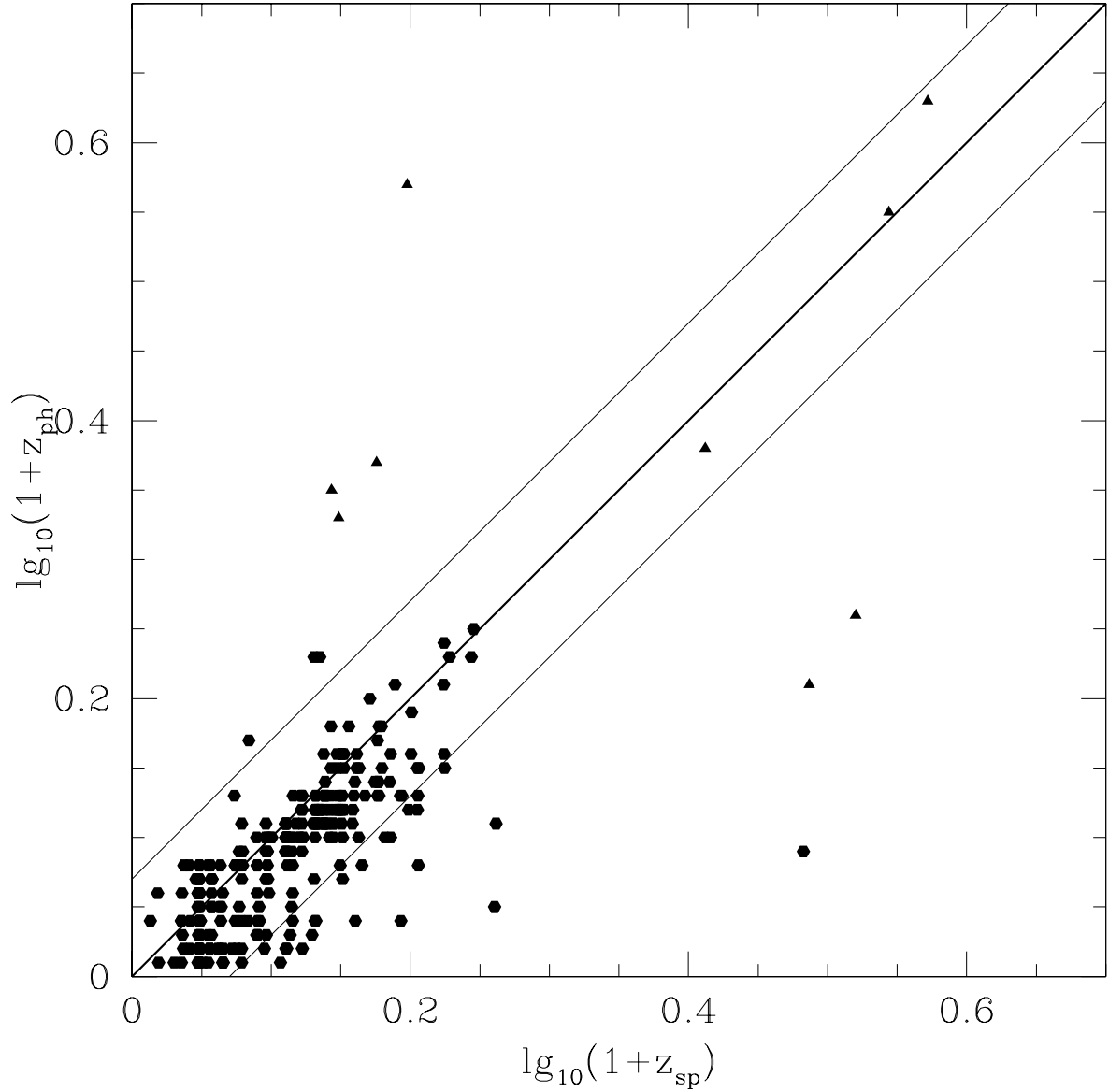}
\caption{Photometric redshifts for the WIYN survey using Ugri  
data only, versus spectroscopic redshift.
Filled circles: galaxies, filled triangles: QSOs.
}
\end{figure}

\begin{figure}
\plotone{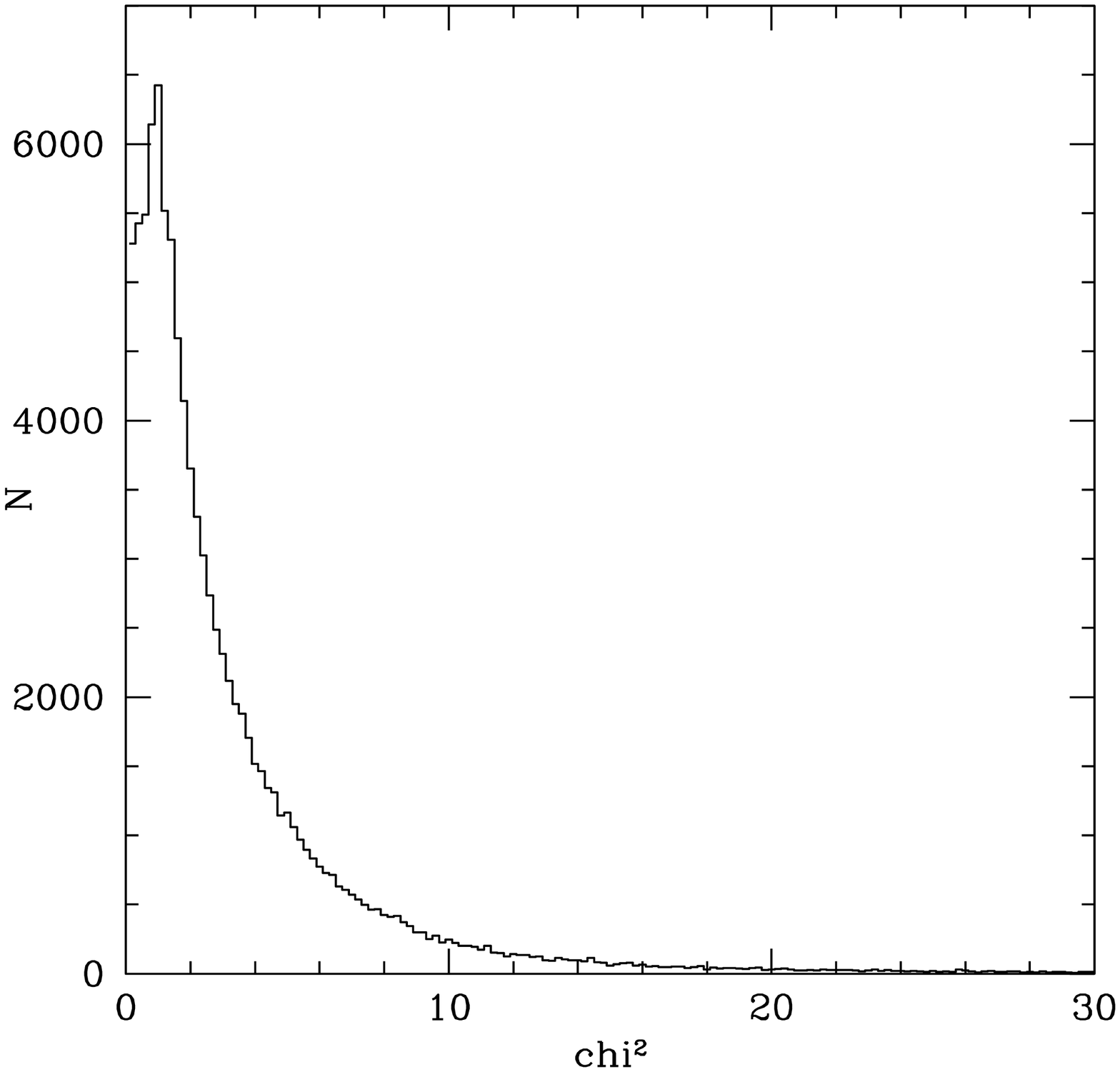}
\caption{Histogram of reduced $\chi^2$ for the photometric redshift fits.
}
\end{figure}

\section{Redshift distributions}
Figure 15 shows the redshift distributions derived in this way for SWIRE-N1 sources with S(3.6) $>$
40 $\mu$Jy, above which flux the optical identifications (to r $\sim$ 23.5) are, according to Fig 9, relatively complete,
with a breakdown into elliptical, spiral + starburst and quasar seds based on the photometric redshift fits.  
For comparison we show in the top panel the predictions
of the Rowan-Robinson (2001), Xu et al (2003) and Pozzi et al (2004) models.  Ellipticals cut off sharply at around z $\sim$ 1.0, and spirals
at z $\sim$ 1.3

Figure 16 shows the redshift distribution for Lockman-VF sources with S(3.6) $>$ 10 $\mu$Jy, for
which the optical identifications (to r $\sim$ 25) are relatively complete.  In the top panel we
show predictions for the same 3 models for $S(3.6) > 10 \mu Jy$.  Again the ellipticals cut off at
z $\sim$ 1.4, but a tail of star-forming galaxies is seen to z $\sim$ 3.

The photometric estimates of redshift for AGN are more uncertain than those for galaxies, due to aliassing problems
and to the effect of variability on photometry,
but the code is effective at identifying Type 1 AGN from the optical and near ir data.  
For some quasars there is significant torus dust emission in the 3.6 and 4.5 $\mu$m bands,
so inclusion of these bands in photometric redshift determination can make the fit worse rather than better.
We have therefore omitted the 3.6 and 4.5 $\mu$m bands if S(3.6)/S(r) $>$ 4.
Note that only
5 $\%$ of SWIRE sources are identified by the photometric redshift code as Type 1 AGN.  

Extrapolating from Figs 9 and 10, we estimate that no more than 5-10 $\%$ of sources are missing from Figs 15 and 16 due to the optical
incompleteness of these two surveys.
However the actual proportions of sources which are either optically blank or have too few
optical bands for photometric redshift determination is much higher than this, 25-35 $\%$.
Further investigation is needed to confirm the reality of these sources and their nature.
Potentially they could substantially shift the shape of the redshift distribution towards higher redshifts.
Further discussion of blank field sources is given in section 7.

\begin{figure}
\plotone{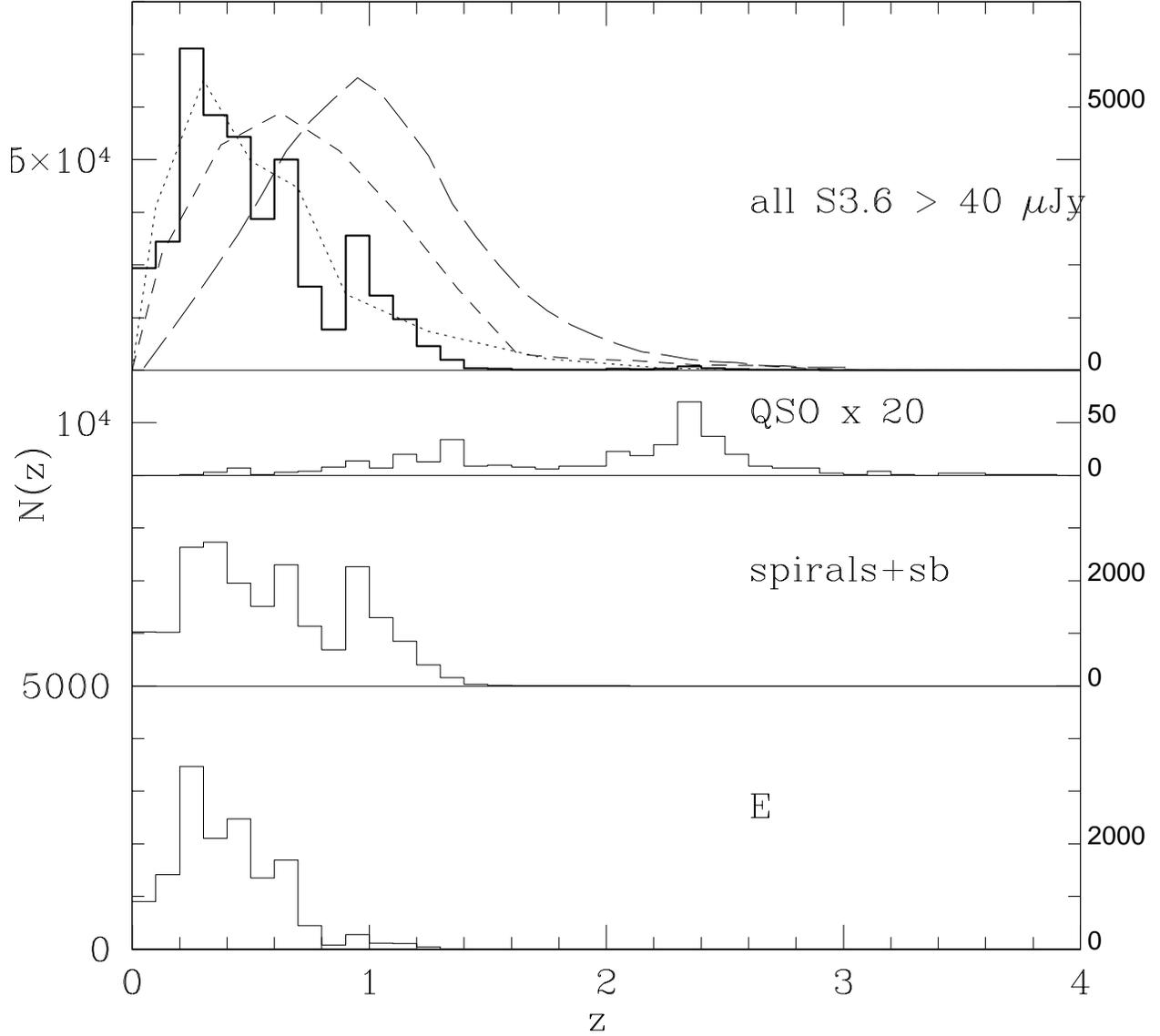}
\caption{Photometric redshift histogram for SWIRE ELAIS-N1 sources with good 4-band optical IDs, and S(3.6) $>$ 40 $\mu$Jy.  
Top panel: all sources, dotted curve: prediction of Rowan-Robinson (2001),
short-dashed curve: prediction of Pozzi et al (2004),
long-dashed curve: prediction of Xu et al (2003, model E);
lower panels: separate breakdown of contributions of ellipticals, spirals
(Sab+Sbc+Scd), starbursts (Sdm+sb), and quasars.  The histogram for quasars has
been multiplied by 20.
}
\end{figure}

\begin{figure}
\plotone{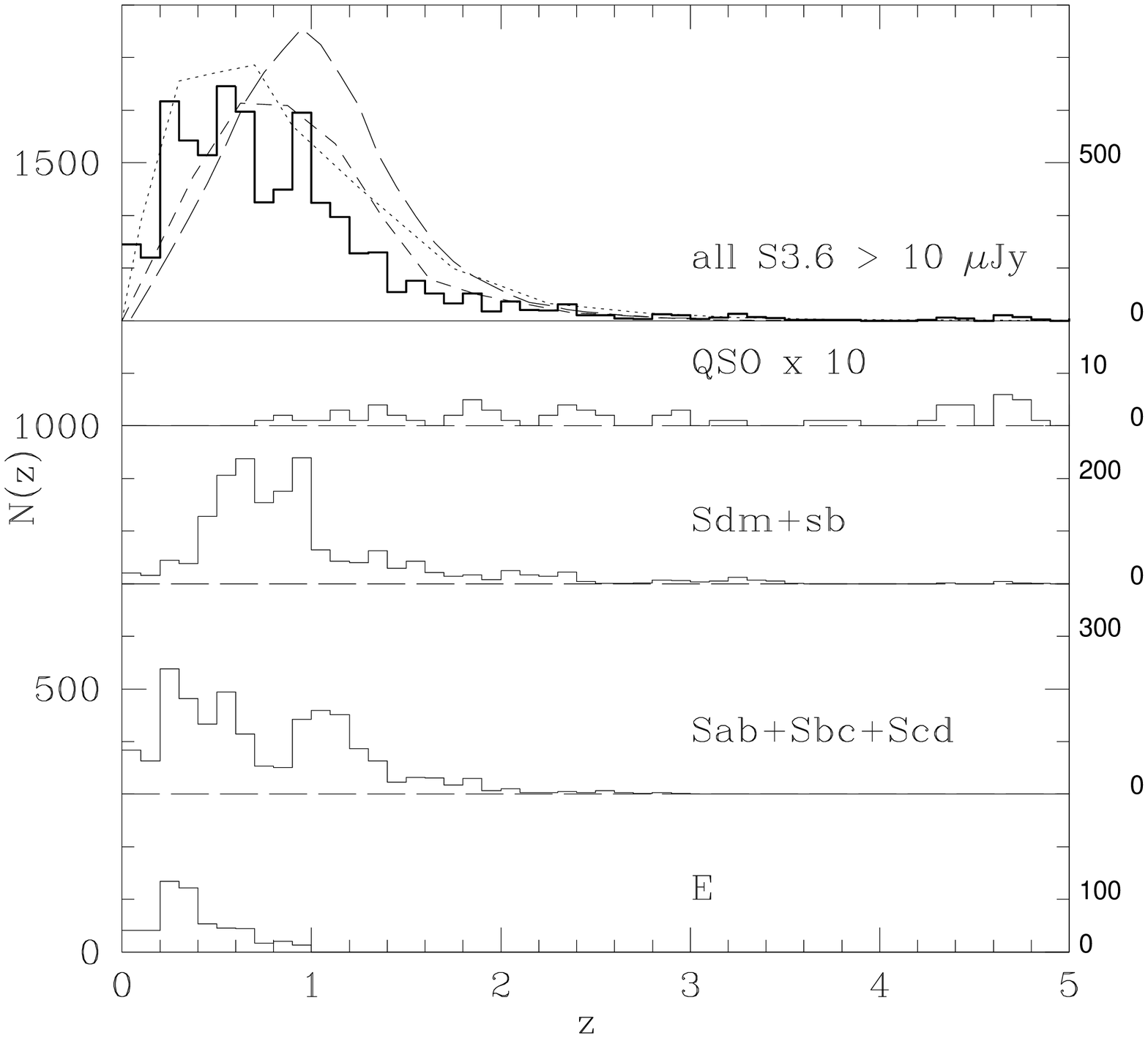}
\caption{Redshift histogram for SWIRE Lockman-VF sources with good 4-band optical IDs, and  S(3.6) $>$ 10 $\mu Jy$.
Top panel: all sources, dotted curve: prediction of 
Rowan-Robinson (2001), short-dashed curve: prediction of Pozzi et al (2004),
long-dashed curve: prediction of Xu et al (2003, model E);
lower panels: separate breakdown of contributions of ellipticals, spirals
(Sab+Sbc+Scd), starbursts (Sdm+sb), and quasars.  The histogram for quasars has
been multiplied by 10. 
}
\end{figure}

\section{Bolometric infrared and optical luminosities}
For sources detected at 70 or 160 $\mu$m, or with an infrared excess at 4.5-24 $\mu$m, relative to the template used for photometric redshift fitting, 
in at least two bands (one of which we require to be 8 or 24 $\mu$m),
we have determined the best-fitting out of cirrus, M82 starburst, Arp220 starburst or AGN dust torus 
infrared templates (as defined in section 3). From the experience of modeling IRAS sources
(eg Rowan-Robinson and Crawford 1989), confirmed by the sed modeling described in section 3,
we allow (a) a mixture of cirrus and M82 templates, (b) a mixture of AGN dust torus and M82 templates, or
(c) an Arp 220 template (using a mixture of AGN dust torus plus A220 template did not improve the $\chi^2$ distribution).
We classify all galaxies with a parameter $n_{sed}$ which is 1 for for an sed dominated by cirrus,
2 for sed dominated by M82 starburst, 3 for Arp 220 starburst, 4 for sed dominated by an AGN dust torus,
5 if there is an infrared excess in only a single band (in this case we can not say what the appropriate
infrared template is), and 6 for no infrared excess (bulge or elliptical dominated, or simply too faint
for dust emission to be detected).  Several figures in this paper have been colour-coded using this
parameter.  
Table 2 gives the break down of the N1 sources with photometric redshift by their
mid-ir (4.5-24 $\mu$m) template type.  The proportion of cirrus galaxies (31$\%$) agrees well
with the figure given by Yan et al (2004) for the Spitzer First Look Survey (30$\%$).  Note the
high proportion of galaxies whose mid-ir seds are fitted by an AGN dust torus template (29$\%$).
Only 8 $\%$ of these are Type 1 AGN according to the optical-nir template fitting.  25$\%$ are fitted
with galaxy templates in the optical-nir and have $L_{ir} > L_{opt}$ and so have to be Type 2 AGN.
The remainder have $L_{ir} < L_{opt}$ so can be Seyferts, in which the optical AGN fails to be
detected against the light of the host galaxy. The implied dust covering factor, $\geq 75\%$, is 
much higher than that inferred for bright optically selected quasars, $\sim 30\%$ (Rowan-Robinson 1995).  

We can estimate the bolometric luminosity corresponding to the infrared template 
and to the optical template used for photometric redshift determination.
Figure 17 shows the ratio of bolometric infrared to optical luminosity, $lg_{10}(L_{ir}/L_{opt})$, versus bolometric infrared luminosity,
 $lg_{10} L_{ir}$
for 916 galaxies detected in N1 at either 70 or 160 $\mu$m (or both), with different symbols/colours for the
different infrared sed types.
For star-forming galaxies the parameter $L_{ir}/L_{opt}$ can be interpreted as approximately $10^{-9} \dot{M_*}/M_* (yr^{-1}$,
since $L_{ir} \sim 10^{10} \dot{M_*}$ and $L_{opt} \sim 10 M_*$,
i.e. as $10^{-9} \tau^{-1}$, where $\tau$ is the time-scale in yrs to accumulate the present stellar mass, forming
stars at the current rate ($\dot{M_*}/M_* = \tau^{-1}$ is the specific star formation rate).
For galaxies with $L_{ir} < 10^{10} L_{\odot}$, a significant fraction of the infrared emission is
reemission of starlight absorbed by (optically thin) interstellar dust, so $L_{ir}/L_{opt}$ should
be interpreted as the optical depth of the interstellar dust.  Many very low values of $L_{ir}/L_{opt}$
($<$ 0.2) are due to elliptical galaxies with a small amount of star-formation.
Fig 18 shows the same plot for 2913 galaxies based on 4.5-24 $\mu$m data from a 0.8 sq deg sub-area of N1 (tile 2-2).  
These figures contains a wealth of information.  The cirrus galaxies at
low infrared luminosities ($< 10^{11} L_{\odot}$) typically have values of $lg_{10}(L_{ir}/L_{opt})$ of -0.2,
in agreement with the values found for nearby IRAS galaxies (Rowan-Robinson et al 1986).  However there
is the interesting population of luminous and ultraluminous cirrus galaxies, with $L_{ir} > L_{opt}$, discussed in
section 3.  
7 $\%$ of the cirrus galaxies identified in the N1 area have $L_{ir} > 3.10^{11} L_{\odot}$ and 2 $\%$ have
$L_{ir} > 10^{12} L_{\odot}$.  The implications are that (1) the quiescent phase of star formation was significantly
more luminous in the past (as assumed in the count models of Rowan-Robinson 2001), (2) the dust opacity of
the interstellar medium in galaxies was higher  at z $\sim$ 1, as expected from galaxy models with
star-formation histories that peak at z = 1-2 (Pei et al 1999, Calzetta and Heckaman 1999, Rowan-Robinson 2003).
The galaxies with Arp 220 templates tend to have high values of $L_{ir}/L_{opt}$, consistent
with the idea that they are very dusty.  M82 type starbursts are less extreme and range over rather similar 
values of $L_{ir}$ and  $L_{ir}/L_{opt}$  to the cirrus galaxies.
Figure 19 shows a histogram of reduced $\chi^2$ for the infrared sed template fitting, for an assumed
typical error of 0.06 dex (corresponding to $\pm 15 \%$), which gives a measure of the goodness of fit of the ir templates.

For galaxies detected at 70 or 160 $\mu$m we can compare the infrared template fits and bolometric luminosity estimates derived from
4.5-160 $\mu$m data, and from 4.5-24 $\mu$m only.  For cirrus galaxies the agreement is good but for other types it
is not so good and for all sources the rms value $lg_{10}(L_{bol,4.5-160}/L_{bol,4.5-24})$ is 0.30, corresponding to a factor 2
uncertainty in $L_{bol}$ estimated from 4.5-24 $\mu$m only.  This is unfortunate because we have 70 or 160 $\mu$m data for
only a small percentage of the SWIRE galaxies with photometric redshifts.

To summarise (see Table 2), for 63 $\%$ of the 126193 SWIRE N1 galaxies with photometric redshifts, we have no information at
all on the dust emission (no significant infrared excess relative to the starlight/QSO models), For 36 $\%$ we have an 
approximate estimate based on 4.5-24 $\mu$m data which gives the bolometric luminosity only to an accuracy of a factor of 2,
and for just 0.7 $\%$ of galaxies do we have a more accurate estimate based on longer wavelength data.

\begin{figure}
\plotone{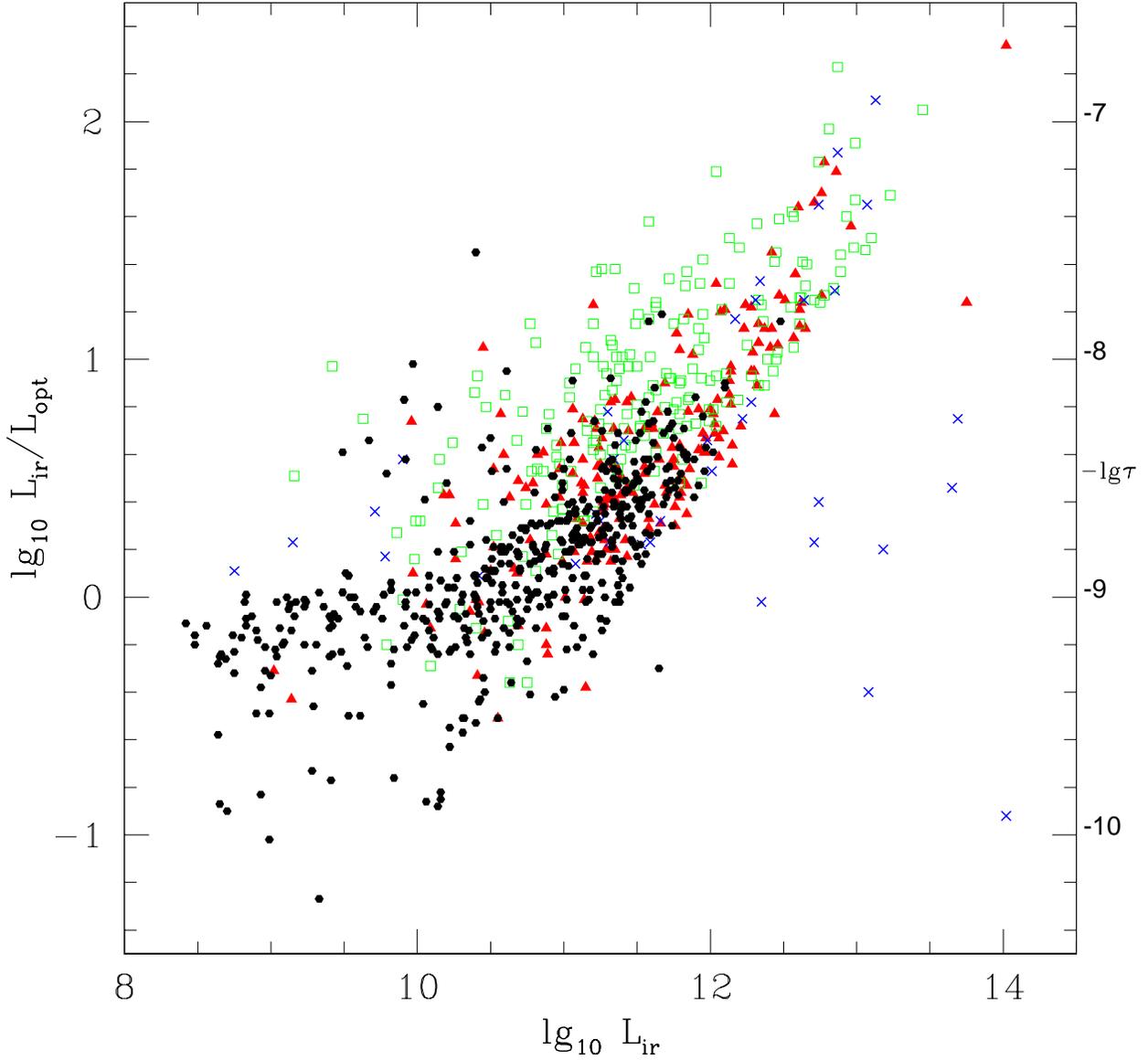}
\caption{Ratio of bolometric infrared to optical luminosity, $lg_{10}(L_{ir}/L_{opt})$, versus bolometric infrared luminosity,
$lg_{10} L_{ir}$,
for all sources in N1 detected at either 70 or 160 $\mu$m.
Filled circles (black): galaxies with cirrus seds, filled triangles (red): M82 starbursts, open squares (green): A220 starbursts,
crosses (blue): AGN dust tori.
Right-hand scale gives values of the specific star-formation rate, $lg_{10}(\dot{M_*}/M_*), (yrs^{-1})$.
}
\end{figure}

\begin{figure}
\plotone{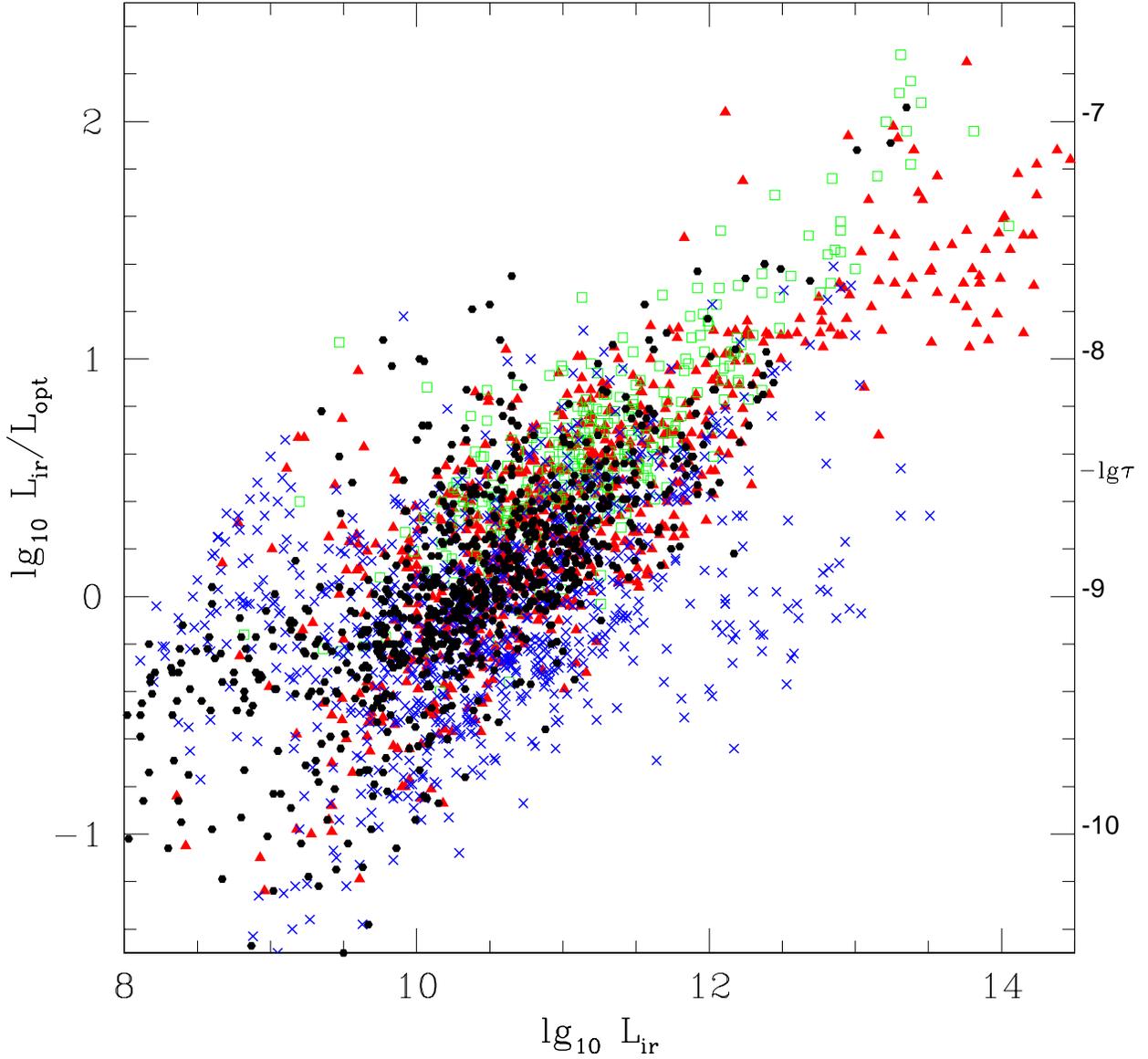}
\caption{Ratio of bolometric infrared to optical luminosity, $lg_{10}(L_{ir}/L_{opt})$, versus bolometric infrared luminosity,
$lg_{10} L_{ir}$,
for SWIRE-N1 (tile 2-2) sources with photometric redshifts and  at least 2 bands at $\lambda > 4.5 \mu m$
with excess infrared emission above the optical galaxy template.
Filled circles (black): galaxies with cirrus seds, filled triangles (red): M82 starbursts, open squares (green): A220 starbursts,
crosses (blue): AGN dust tori.
Right-hand scale gives values of $lg_{10}(\dot{M_*}/M_*), (yrs^{-1})$.
}
\end{figure}

\begin{figure}
\plotone{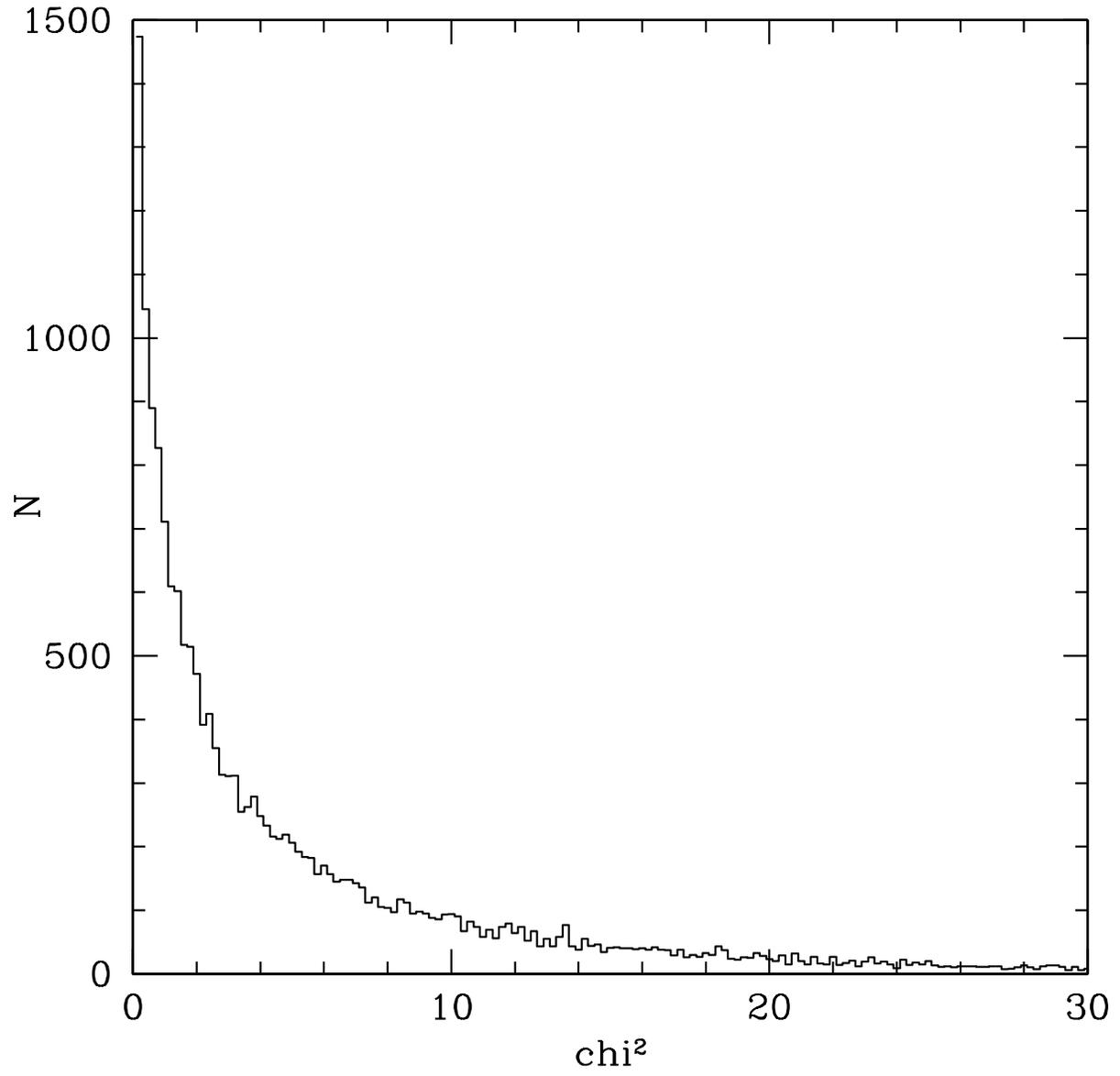}
\caption{Histogram of reduced $\chi^2$ for the ir template fits, assuming typical error of 0.06 dex.
}
\end{figure}

The code of Rowan-Robinson (2003) imposes an upper limit on the B band luminosity for starlight, which translates
approximately to a limit on the optical bolometric luminosity (and stellar mass),  
which manifests itself as the sloping cutoff to the right of Fig 17.  To the right of this
limit can be seen the Type 1 quasars, some of which are best fitted by AGN dust tori, but others prefer
M82 starbursts.
While type 1 AGN would be expected to lie below the $L_{ir} = L_{opt}$ line, 
since only part of the optical light we see is being intercepted by the dust torus, many of the galaxies whose
mid and far infrared seds are dominated by AGN dust tori have  $L_{ir} > L_{opt}$, consistent with
being Type 2 AGN.
Further discussion of SWIRE quasars is given by Hatziminaoglou et al (2004).

Quite a few objects are seen to have extremely high infrared luminosities, in the hyperluminous class
($> 10^{13} L_{\odot}$), most of them associated with quasars at high redshifts.  Clearly spectroscopy
is needed to determine these redshifts more accurately.

Since the sources plotted in Fig 18 are mostly detected at 24 $\mu$m it is of interest to
consider whether PAH features are affecting the detectability of 24 $\mu$m sources.
Figure 20 shows redshift distributions for 24 $\mu$m  broken down by infrared sed type.  
Cirrus sources, and M82 starburst show a strong cutoff at z $\sim$ 1.4, which
may be partly due to the onset of the 10 $\mu$m silicate absorption feature.
For Arp 220 sources, the cutoff starts at lower redshift ($\sim$ 1), consistent with their
deeper and broader silicate absorption.  
A secondary peak for M82 starbursts (and possibly cirrus sources) at z = 0.8-1.4 can be interpreted as sources having their detectability at 24 $\mu$m enhanced by the 10-12 $\mu$m PAH 
feature.   Sources with an infrared excess in a single band have a broad distribution 
between z = 0.2 - 1.4, and are consistent with being M82 starbursts.
The optically blank 24 $\mu$m sources would be likely to shift this distribution to higher
redshifts, as predicted for example by the model of Rowan-Robinson (2001) shown.
Chary et al (2004) give a predicted redshift distribution for the deeper (20 $\mu$Jy) GOODS-SV survey,
based on a fit to the 24 $\mu$m counts, which has a broad peak at z = 0.8-2.

Yan et al (2004) have discussed the nature of SPITZER 24 $\mu$m sources with
reference a plot of S(24)/S(8) versus S(24)/S(r) and drawn attention to sources
with high values of S(24)/S(r) as a potential new population.  
Specifically they find that 23 $\%$ of the FLS sample have S(24)/S(r) $>$ 300 and suggest these are
probably luminous infrared galaxies at z $>$ 1.  They find that 2$\%$ of their sample have S(24)/S(r)
$>$ 1000 and speculate that these may by obscured AGN at z $>$ 0.6.
In Fig 21 we show S(24)/S(r) versus redshift for the Lockman-VF with different symbols/colours for
different sed types.  For the objects plotted here, no new sed types are required.
Galaxies with an infrared excess only at 24 $\mu$m (open yellow triangles) and
z $>$ 1.5 are Type 1 AGN, according to their optical seds.  Many of those at
z $<$ 1.5 could be Type 2 AGN.  Figure 10 suggests that blank field sources
(r $>$ 25) are mainly AGN dust tori or M82 starbursts.  A 24 $\mu$m optically blank source
with S24 $>$ 70 $\mu$Jy would have S(24)/S(r) $>$ 200, so from Fig 21 would be
likely to have z $>$ 0.5. Whether the 24 $\mu$m sources
without optical counterparts represent a new population, or simply slightly more
distant counterparts of the identified sources, remains to be seen.  It is worth noting that
S(24)/S(r) acts as a crude redshift indicator,
in that if S(24)/S(r) $<$ 5, then z $<$ 0.5.  However higher values can show a very
wide range of z.

\begin{figure}
\plotone{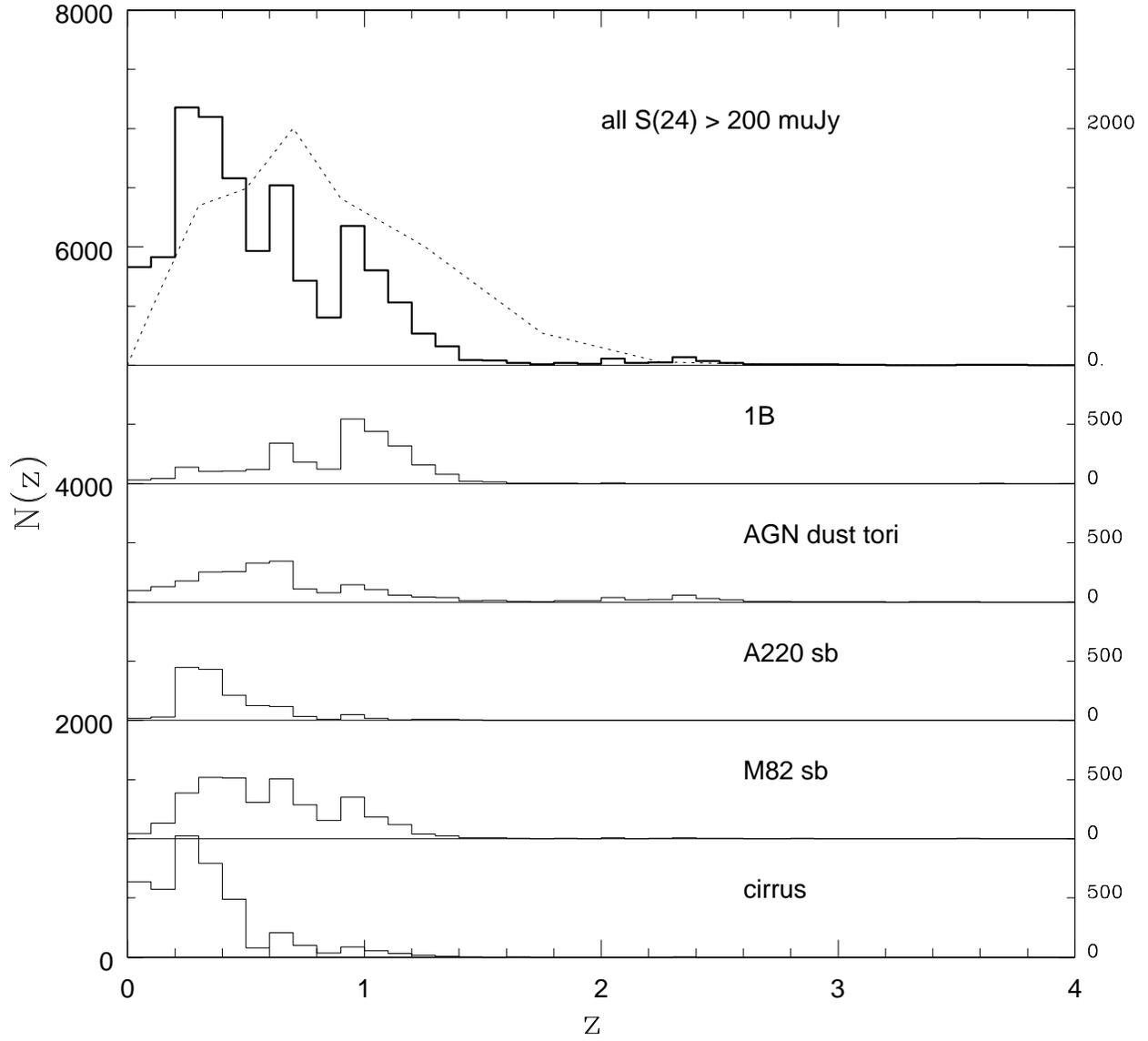}
\caption{Redshift histogram for 24 $\mu$m sources brighter than 200 $\mu Jy$ broken up by
infrared sed type.
Dotted curve: prediction of Rowan-Robinson (2001).
}
\end{figure}

\begin{figure}
\plotone{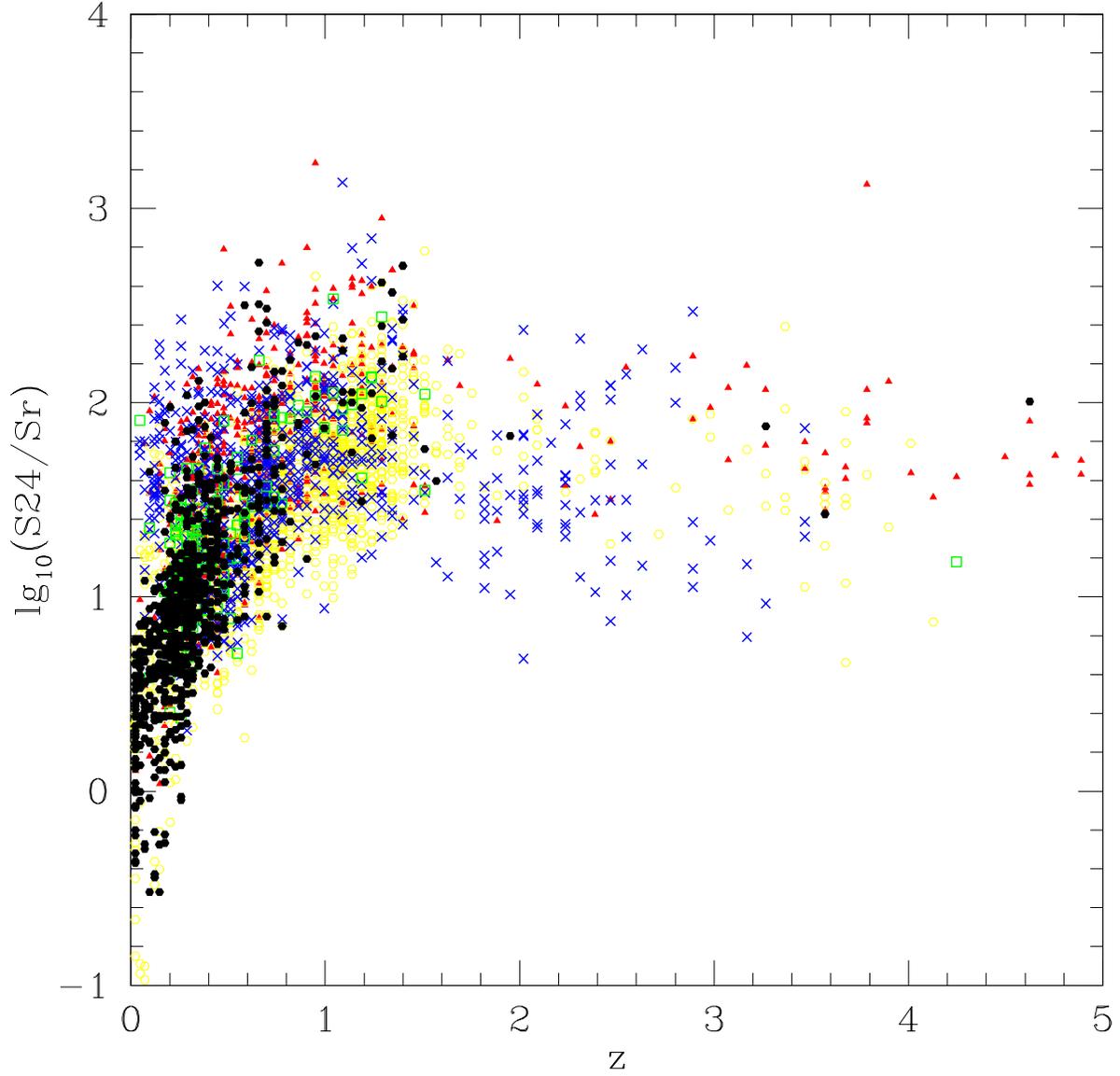}
\caption{ $\log_{10}S(24)/S(r)$ for galaxies and quasars in Lockman-VF,
with different symbols for
infrared template types:
filled circles (black): cirrus, filled triangles (red): M82 starburst, open squares (green): Arp220,
crosses (blue): AGN dust tori, open circles (yellow): only one band with infrared excess.
}
\end{figure}

\section{Conclusions}

We have discussed optical associations, spectral energy distributions and photometric redshifts
for SWIRE sources in the ELAIS-N1 area and the Lockman Validation Field.
The band-merged IRAC (3.6, 4.5, 5.8 and 8.0 $\mu$m) and MIPS (24, 70 and 160 $\mu$m) data
have been associated with optical UgriZ data from the INT Wide Field Survey in 
ELAIS-N1, and with our own optical Ugri data in Lockman-VF.  Criteria for eliminating
spurious infrared sources and for carrying out star-quasar-galaxy separation,
and statistics of the identification rate, are given.

The spectral energy distributions of selected SWIRE sources in ELAIS-N1
with spectroscopic redshifts,
are modelled in terms of a simple set of galaxy and quasar templates in the optical 
and near infrared, and with a set of dust emission templates (cirrus, M82 starburst,
Arp 220 starburst, and AGN dust torus) in the mid and far infrared.

The optical data, together with the IRAC 3.6 and 4.5 $\mu$m data, have been used
to determine photometric redshifts.  For galaxies with known spectroscopic redshifts
there is a notable improvement in the photometric redshift when the IRAC data are used,
with a reduction in the rms scatter from 10$\%$ in (1+z) to 7$\%$.  While further spectroscopic
data are needed to confirm this result, the prospect of determining good photometric 
redshifts for much of the SWIRE survey, expected to yield over 2 million extragalactic 
objects, is excellent.  Some modifications to the optical-nir templates were required in
the previously uninvestigated wavelength region 2-5 $\mu$m.

The photometric redshifts are used to derive the 3.6 $\mu$m redshift distribution.
Agreement with model predictions are quite good.
For those sources with a clear mid infrared excess, relative to the galaxy starlight 
model used for the optical and near infrared, the distribution of the different 
infrared sed types in the $L_{ir}/L_{opt}$ versus $L_{ir}$ plane, where $L_{ir}$ and $L_{opt}$
are the infrared and optical bolometric luminosities, is discussed.
Low luminosity cirrus galaxies have values of $L_{ir}/L_{opt}$ similar to those
found with IRAS data, but there is a new population of luminous cold
cirrus galaxies with $L_{ir} > L_{opt}$, which implies substantial dust
extinction.  These appear to present an enhanced rate of quiescent star formation
compared to those seen in spirals in our neighbourhood.

There is a surprisingly high incidence of galaxies with Arp220-like seds (10$\%$), which was
not predicted by the counts models of Rowan-Robinson (2001).  They tend to have
high values of $L_{ir}/L_{opt}$.  
There is also a high proportion of galaxies whose mid-ir seds are fitted by an AGN dust torus template (29$\%$).
Of these only 8 $\%$ of these are Type 1 AGN according to the optical-nir template fitting
while 25$\%$ are fitted
with galaxy templates in the optical-nir and have $L_{ir} > L_{opt}$ and so have to be Type 2 AGN.
The remainder have $L_{ir} < L_{opt}$ so can be Seyferts, in which the optical AGN fails to be
detected against the light of the host galaxy. The implied dust covering factor, $\geq 75\%$, is 
much higher than that inferred for bright optically selected quasars.

For galaxies detected at 70 or 160 $\mu$m we have compared the infrared template fits and bolometric luminosity estimates derived from
4.5-160 $\mu$m data, and from 4.5-24 $\mu$m only.
The rms value $lg_{10}(L_{bol,4.5-160}/L_{bol,4.5-24})$ is 0.30, corresponding to a factor 2
uncertainty in $L_{bol}$ estimated from 4.5-24 $\mu$m only.  This is unfortunate because we have 70 or 160 $\mu$m data for
only 916 of the 126193 SWIRE galaxies with photometric redshifts in N1.
Thus for 60 $\%$ of the SWIRE N1 galaxies with photometric redshifts, we have no information at
all on the dust emission (no significant infrared excess relative to the starlight/QSO models), For 39 $\%$ we have an
approximate estimate based on 4.5-24 $\mu$m data which gives the bolometric luminosity only to an accuracy of a factor of 2,
and for just 0.7 $\%$ of galaxies do we have a more accurate estimate based on longer wavelength data.
Future instruments like PACS on Herschel, or large cooled apertures like SAFIR and SPICA, will be needed
to fully quantify far infrared emission from galaxies at z $>$ 1.

\section{Acknowledgements}
This work has been supported by the European Network 'POE', HPRN-CT-2000-00138.  We thank the referee
for constructive comments which allowed the paper to be improved.

\begin{table*}
\caption{Properties of SWIRE-ELAIS galaxies with plotted seds.  The three sections of the
table correspond to the galaxies of Figs 4-6, respectively, in order from the bottom of
the figures. $n_{sed}$ is the infrared template type, and $n_{typ}$ is the optical template type
(see text).  Bracketed redshifts are photometric, rest are spectroscopic.}
\begin{tabular}{crrrrrrrrr}
 & & & & & & & & &\\
SWIRE RA & dec & ELAIS name & z & $n_{sed}$ & $lg L_{ir}$ & C & $n_{typ}$ & $A_V$ & $lg L_{opt}$ \\
 & (2000) & & & & ($L_o$) & & & ($L_o$) \\
&&&&&&&&&\\
241.14116 & 54.74209 & 160433.8+544432 & 0.0751 & 1 & 10.83 & 0 & 2 & 0 & 10.64\\
241.47246 & 54.37391 & 150553.3+542225 & 0.2116 & 1 & 11.23 & 2 & 2 & 0 & 11.29\\
241.90221 & 53.95853 & 160736.5+535731 & 0.0298 & 1 & 10.08 & 5 & 3 & 0 & 10.08\\
241.90823 & 54.76723 & 160737.9+544601 & 0.0913 & 1 & 10.94 & 7 & 2 & 0 & 10.75\\
242.01546 & 54.88388 & 160803.8+543202 & 0.0528 & 1 & 10.83 & 10 & 2 & 0 & 10.83\\
242.05882 & 54.47658 & 160814.2+542836 & 0.1198 & 1 & 10.87 & 12 & 3 & 0 & 10.36\\
242.14723 & 53.83896 & 160835.2+535022 & 0.0627 & 1 & 10.48 & 14 & 1 & 0 & 10.27\\
242.21648 & 54.21648 & 160937.5+541258 & 0.0862 & 1 & 10.74 & 16 & 1 & 0 & 10.42\\
242.76556 & 54.72275 & 161103.7+544322 & 0.0627 & 1 & 10.33 & 18 & 3 & 0 & 10.34\\
243.28314 & 54.86163 & 161308.1+545141 & 0.0700 & 1 & 10.38 & 20 & 3 & 0 & 10.42\\
&&&&&&&&&\\
241.67583 & 55.83316 & - & (0.479) & 1 & 12.38 & -3 & 4 & 0 & 11.42 \\
242.62575 & 55.02607 & 161030.1+550135 & 0.2740 & 1 & 11.78 & 0 & 1 & 0 & 11.19\\
244.18948 & 54.51995 & 161645.3+543111 & 0.2221 & 1 & 11.40 & 2 & 3 & 0 & 11.80\\
240.84489 & 54.71045 & 160322.8+544237 & 0.2160 & 2 & 11.32 & 5 & 3 & 0 & 10.78\\
241.58780 & 53.72672 & 160621.0+534336 & 0.2067 & 2 & 11.26 & 7 & 2 & 0 & 11.13\\
241.92168 & 55.03159 & 160741.1+550152 & 0.4599 & 2 & 12.02 & 9 & 2 & 0 & 10.66\\
242.10521 & 54.63612 & 160825+543809   & 0.9070  & 2 & 12.45 & 11 & 4 & 0 & 11.40\\
242.23936 & 54.17452 & 160857.6+541027 & 0.2687 & 2 & 11.15 & 14 & 1 & 0 & 10.88\\
243.74185 & 54.14808 & 161457.9+540853 & 0.2150 & 2 & 11.27 & 16 & 4 & 0 & 10.93\\
243.96436 & 54.25998 & 161551.3+541536 & 0.2148 & 2 & 11.27 & 18 & 4 & 0 & 10.93\\
&&&&&&&&&\\
241.39261 & 54.69358 & 160534.3+544136 & (0.148) & 3 & 10.90 & -5 & 1 & 0 & 10.44\\
242.20854 & 54.73938 & 160850+544422   & (0.514) & 3 & 11.96 & -3 & 2 & 0 & 11.32\\
242.25496 & 54.30231 & 160901+541808 & 0.3320 & 3 & 11.50 & 0 & 4 & 1.0 & 10.80\\
242.52415 & 54.17478 & 161005.8+541029 & 0.0636 & 3 & 10.63 & 3 & 2 & 0 & 10.84\\
242.58041 & 54.36514 & 161019+542153   & 0.2074 & 3 & 11.22 & 6 & 4 & 0.7 & 10.45\\
241.73061 & 53.67138 & 160655.5+534016 & 0.2136 & 4 & 11.13 & 10 & 7 & 0 & 11.19\\
242.30492 & 53.90829 & 160913+535429 & 0.9924 & 4 & 12.14 & 11 & 7 & 0 & 12.48\\
243.15944 & 53.38187 & - & 2.138 & 4 &  12.98 & 12 & 7 & 0 & 13.48\\
241.77142 & 53.59966 & - & 3.653 & 4 & 13.80 & 13 & 7 & 0 & 13.75\\
240.75597 & 54.75597 & - & 0.7280 & 4 & 12.54 & 17 & 7 & 0 & 12.29\\

\end{tabular}
\end{table*}

\begin{table}
\caption{Statistics of SWIRE Catalogue in EN1 and Lockman-VF}
\begin{tabular}{crr}
&&\\
area & 6.5 & 0.3 sq deg\\
total SWIRE Catalogue & 301021$^a$  & 19140$^a$\\
optically blank & 96155$^b$ & 2991\\
optical ID but no r, or not 4 bands & 32721 & 2331\\
(out of 3.6, 4.5 $\mu$m, UgriZ) & \\
stars & 38500 & 1154\\
phot z failed & 7452 & 966\\
galaxies with phot z & 126193 & 11698\\
& & \\
of which 4.5 $\mu$m & 101139 & 8931\\
5.8 $\mu$m & 27554 & 2109\\
8.0 $\mu$m & 27074 & 1472\\
24.0 $\mu$m & 38601 & 878\\
70 $\mu$m & 870 & 22\\
160 $\mu$m & 406 & 14\\
& & \\
 E & 29383 (23$\%$) & 2361 (20$\%$)\\
 Sab & 19326 (15$\%$) & 1292 (11$\%$)\\
 Sbc & 13483 (11$\%$) & 1636 (14$\%$)\\
 Scd & 29235 (23$\%$) 3209 (28$\$$)\\
 Sdm & 9512 (7$\%$) & 949 (8$\%$)\\
 sb & 19095 (15$\%$) & 1769 (15$\%$)\\
 QSO & 6159 (5$\%$) & 481 (4$\%$)\\
 & &\\
 cirrus & 8091 (31$\%$) & 177 (25$\%$)\\
 M82 & 7595 (29$\%$) & 166 (24$\%$)\\
 A220 & 2624 (10$\%$) & 165 (24$\%$)\\
 AGN dust torus & 7598 (29$\%$) & 197 (28$\%$)\\
 single band ir excess & 20793 & 749\\
 no ir excess & 79492 & 10244\\
 & & \\
$^a$ all detected at 3.6 $\mu$m &\\
$^b$ includes small number of &\\
sources in WFS chip gaps &\\
\end{tabular}
\end{table}






\end{document}